\documentclass[3p, 12pt, sort&compress]{elsarticle}

\usepackage{soul}

\pdfoutput = 1

\usepackage[font={it}]{caption}

\usepackage{dsfont, 
  amssymb, amsmath,
  graphicx,
  hyperref,
  color,
  braket,
  array,
  mathrsfs,
  subfig,
  upgreek,
  slashbox}

\usepackage{bm}

\usepackage{slashed}

\numberwithin{equation}{section}

\journal{Nuclear Physics B}

\makeatletter

\def\ps@pprintTitle{
 \let\@oddhead\@empty
 \let\@evenhead\@empty
 \def\@oddfoot{}
 \let\@evenfoot\@oddfoot}

\long\def\MaketitleBox{
  \resetTitleCounters
  \def\baselinestretch{1}
  \begin{center}
   \def\baselinestretch{1}
    \Large\@title\par\vskip18pt
    \normalsize\elsauthors\par\vskip10pt
    \footnotesize\itshape\elsaddress\par\vskip36pt
    \rule{\textwidth}{1.5pt}\vskip12pt 
    \ifvoid\absbox\else\unvbox\absbox\par\vskip10pt\fi
    \ifvoid\keybox\else\unvbox\keybox\par\vskip10pt\fi
    \rule{\textwidth}{1.5pt}\vskip12pt 
    \end{center}
  }

\renewcommand\subsection{\@startsection{subsection}{2}{\z@}
           {18\p@ \@plus 6\p@ \@minus 3\p@}
           {9\p@ \@plus 6\p@ \@minus 3\p@}
           {\normalfont\normalsize\itshape\bfseries}}

\gdef\emailauthor#1#2{\stepcounter{ead}
     \g@addto@macro\@elseads{\raggedright
      \let\corref\@gobble
      \eadsep\newline\texttt{#1} (#2)\def\eadsep{\unskip,\space}}
}

\def\appendixname{Appendix}
\renewcommand\@makefntext[1]{#1}

\makeatother

\newcommand{\Del}[4]{[\Delta^N_{#1}(#2)]_{#3}^{\phantom #3 #4}}
\newcommand{\gC}{\widetilde{C}}
\newcommand{\del}[4]{[\Delta_{#1}^{#2}]_{#3}^{\phantom #3 #4}}
\newcommand{\delh}[4]{[\widehat{\Delta}_{#1}^{#2}]_{#3 #4}}

\newcommand{\CP}{C\!P}
\newcommand{\gCP}{\widetilde{C}\!P}

\newcommand{\ve}[1]{\mathbf{#1}}
\DeclareMathOperator{\Tr}{Tr}

\newcommand{\Esdu}[4]
{\left[(2 E_{#1}(\ve #2))^{-{1}/{2}}\right]_{#3}^{\phantom{#3}#4}}

\newcommand{\mat}[1]{\bm{#1}}

\newcommand{\edu}[4]{\big[e^{#1 i #2 \cdot x}\big]_{#3}^{\phantom{#3} #4}}

\newcommand{\h}[2]{h_{#1}^{\phantom{#1}#2}}
\newcommand{\hs}[2]{h_{\phantom{#1}#2}^{#1}}
\newcommand{\hr}[2]{\mathbf{h}_{#1}^{\phantom{#1}#2}}
\newcommand{\hrs}[2]{\mathbf{h}_{\phantom{#1}#2}^{#1}}
\newcommand{\hrc}[2]{[\mathbf{h}^{\tilde{c}}]^{#1}_{\phantom{#1}#2}}
\newcommand{\hrcs}[2]{[\mathbf{h}^{\tilde{c}}]^{\phantom{#1}#2}_{#1}}

\newcommand{\Tdu}[5]
{{#1}_{#2 \phantom{#3} #4 \phantom{#5}}^{\phantom{#2} #3 \phantom{#4}  #5}}

\DeclareMathAlphabet{\mathpzc}{OT1}{pzc}{m}{it} 

\newcommand{\D}[2]{\mathrm{d}^{#1}{#2}}

\newcommand{\ansatz}{ansatz}

\usepackage[nodayofweek]{datetime} 

\renewcommand{\Re}{\mathrm{Re}}
\renewcommand{\Im}{\mathrm{Im}}

\begin{document}

\begin{frontmatter}
\begin{flushright}
\begin{footnotesize}
 MAN/HEP/2014/13, TUM-HEP-962-14\\
 October 2014
\end{footnotesize} 
\end{flushright}

\title{{\bf {\LARGE Kadanoff--Baym Approach to Flavour Mixing\\[3pt] and Oscillations in Resonant Leptogenesis}}\medskip}

\author[a]{\large P.~S.~Bhupal Dev}

\author[b]{\large Peter Millington}

\author[a]{\large Apostolos Pilaftsis}

\author[a]{\large Daniele Teresi}

\address[a]{Consortium for Fundamental Physics,
  School of Physics and Astronomy, \\ 
  University of Manchester, Manchester M13 9PL, United Kingdom.}

\address[b]{Physik Department T70, James-Franck-Stra\ss e,\\
Technische Universit\"{a}t M\"{u}nchen, 85748 Garching, Germany.}


\begin{abstract}

We describe a loopwise perturbative truncation scheme for quantum transport equations in the Kadanoff--Baym formalism, which does not necessitate the use of the so-called Kadanoff--Baym or quasi-particle ansaetze for dressed propagators. This truncation scheme is used to study flavour effects in the context of Resonant Leptogenesis (RL), showing explicitly that, in the weakly-resonant regime, there exist two \emph{distinct} and pertinent flavour effects in the heavy-neutrino sector: (i) the resonant mixing and (ii) the oscillations between different heavy-neutrino flavours. Moreover, we illustrate that Kadanoff--Baym and quasi-particle ansaetze, whilst appropriate for the flavour-singlet dressed charged-lepton and Higgs propagators of the RL scenario, should not be applied to the dressed heavy-neutrino propagators. The use of these approximations for the latter is shown to capture \emph{only} flavour oscillations, whilst discarding the \emph{separate} phenomenon of flavour mixing.

\end{abstract}

\medskip


\end{frontmatter}

\makeatletter
\def\appendixname{Appendix}
\renewcommand\@makefntext[1]
{\leftskip=0em\hskip1em\@makefnmark\space #1}
\makeatother


\newpage
\begin{small}
\tableofcontents
\end{small}

\section{Introduction}

The scenario of leptogenesis \cite{Fukugita:1986hr} can be regarded as a cosmological consequence of the seesaw mechanism~\cite{seesaw1, seesaw2, seesaw4, seesaw5, seesaw6}, thereby providing an elegant unifying  framework to account for both the observed baryon asymmetry of our Universe (BAU) and the smallness of the light-neutrino masses~\cite{pdg}. The key ingredients are heavy Majorana neutrinos, whose out-of-equilibrium decays provide an initial excess in lepton number ($L$). This excess is subsequently converted to a net baryon number ($B$) via the equilibrated $(B+L)$-violating interactions of the electroweak sphalerons~\cite{Kuzmin:1985mm}. The $L$-violating Majorana mass terms, complex Yukawa couplings and expansion of the Universe fulfill the necessary Sakharov conditions~\cite{Sakharov:1967dj} for dynamically generating the BAU, namely $B$, $C$ and $\CP$ violation, together with out-of-equilibrium dynamics. For a review on various aspects of leptogenesis, see e.g.~\cite{Blanchet:2012bk} and references therein.

An attractive possibility of testing leptogenesis in foreseeable laboratory experiments is provided by the mechanism of Resonant Leptogenesis (RL)~\cite{Pilaftsis:1997dr,  Pilaftsis:1997jf, Pilaftsis:2003gt}. This relies on the fact that the  heavy  Majorana neutrino   self-energy
effects  on  the  leptonic $\CP$-asymmetry (the $\varepsilon$-type effects) become
dominant~\cite{Flanz:1994yx, Covi:1996wh, Buchmuller:1997yu} and get resonantly enhanced, when 
at least two of the heavy  neutrinos have a  small mass difference  
comparable to their  decay  widths~\cite{Pilaftsis:1997dr, Pilaftsis:1997jf}. The resonant enhancement of the $\CP$-asymmetry enables a successful low-scale leptogenesis~\cite{Pilaftsis:2003gt, Pilaftsis:2005rv},  whilst retaining perfect agreement with the active neutrino oscillation data~\cite{pdg}. This level of testability is further improved in the scenario of Resonant $\ell$-Genesis (RL$_{\ell}$), where the final lepton asymmetry is dominantly generated and stored in a {\it single} lepton flavour $\ell$~\cite{Pilaftsis:2004xx, Deppisch:2010fr}. In such models, the heavy neutrinos could be at a scale as low as the electroweak scale~\cite{Pilaftsis:2005rv}, whilst still having sizable couplings to other charged-lepton flavours $\ell'\neq \ell$. Thus, RL$_{\ell}$ scenarios may be testable in the run-II phase of the LHC~\cite{Datta:1993nm,Han:2006ip,Bray:2007ru,Atre:2009rg,Dev:2013wba,Bambhaniya:2014kga} as well as in various low-energy experiments searching for lepton flavour/number violation~\cite{Ilakovac:1994kj,Pilaftsis:2005rv, Deppisch:2010fr,Alonso:2012ji}.

In RL models, with quasi-degenerate heavy Majorana neutrinos, flavour effects in both heavy-neutrino~\cite{Pilaftsis:1997jf, Pilaftsis:1998pd, Endoh:2003mz, Pilaftsis:2005rv, Pilaftsis:2004xx, Deppisch:2010fr, Vives:2005ra, Asaka:2005, Blanchet:2011xq,Ellis:2002eh} and charged-lepton~\cite{Barbieri:1999ma, Pascoli:2006ie,Abada:2006ea, Abada:2006fw,Branco:2006hz,
Nardi:2006fx,   Blanchet:2006be,  De  Simone:2006dd} sectors, as well as the interplay between them, can play an important role in 
determining the final lepton asymmetry. These intrinsically-quantum effects  can be accounted for by extending
the classical  flavour-diagonal Boltzmann equations for the number  densities of individual
flavour species to a  semi-classical evolution equation for a
{\it matrix  of  number densities}, analogous to the formalism presented in~\cite{Sigl:1993} for light neutrinos. This approach, the so-called `density matrix' formalism, has been adopted for various leptogenesis scenarios~\cite{Abada:2006ea, De  Simone:2006dd, Blanchet:2011xq, Akhmedov:1998qx, Asaka:2005, Shaposhnikov:2008pf, Canetti:2012kh, Gagnon:2010kt, Asaka:2011wq, Shuve:2014zua}. A consistent treatment of {\em all} pertinent flavour effects, including flavour mixing, oscillations and off-diagonal (de)coherences, necessitates a {\em fully} flavour-covariant formalism, which was recently developed in~\cite{Dev:2014laa} (for an executive summary, see~\cite{Dev:2014tpa}). This provides a complete and unified description of RL. Moreover, the resonant mixing of different heavy-neutrino flavours and coherent oscillations between them were found to be two {\em distinct} physical phenomena, in analogy with the experimentally-distinguishable phenomena of mixing and oscillations in the neutral $K$, $D$, $B$ and $B_s$-meson systems~\cite{pdg} \footnote{This was also shown explicitly in the context of cascade decays of heavy particles, where the time-scales for mixing and oscillation are well separated~\cite{Boyanovsky:2006yg, Boyanovsky:2014uya, Boyanovsky:2014lqa, Boyanovsky:2014una}.}. In particular, we draw attention to the phenomenon of oscillations via regeneration for the kaon system in medium~\cite{Pais:1955sm,Kabir}. A proper treatment of these flavour effects in this {\em fully} flavour-covariant formalism could lead to a significant enhancement in the final lepton asymmetry, as compared to partially flavour-dependent or flavour-diagonal limits, as illustrated numerically in~\cite{Dev:2014laa} within the context of an RL$_\tau$ model.\footnote{Note that in RL$_\ell$ scenarios, the quantum (de)coherence effects in the charged-lepton sector must also be included, which might further contribute to the enhancement of the final lepton asymmetry, depending upon the model parameters~\cite{Dev:2014laa}.}  

On the other hand, there has been significant progress in the literature, attempting to go beyond the semi-classical `density matrix' approach~\cite{Sigl:1993} to transport phenomena by means of the quantum field-theoretic analogues of the Boltzmann equations (see e.g.~\cite{Prokopec:2003pj, Prokopec:2004ic}), known as the Kadanoff--Baym (KB) equations~\cite{Baym:1961zz} (for reviews, see~\cite{KB, Blaizot:2001nr,Berges:2004yj}). This system of quantum Boltzmann equations is commonly derived from the Cornwall-Jackiw-Tomboulis (CJT) effective action~\cite{Cornwall:1974vz,AmelinoCamelia:1992nc,Berges:2004yj,
Berges:2005hc,Pilaftsis:2013xna} of the Schwinger-Keldysh~\cite{Schwinger:1961, Keldysh:1964} closed-time path (CTP) formalism of non-equilibrium thermal field theory~\cite{Jordan:1986ug, Calzetta:1986ey, Calzetta:1986cq}. The KB equations are manifestly non-Markovian, describing the non-equilibrium time-evolution of two-point correlation functions, and have been studied extensively in various scenarios of leptogenesis~\cite{Buchmuller:2000nd, De Simone:2007rw, De Simone:2007pa, Cirigliano:2007hb, Garny:2009qn,Garny:2009rv, Cirigliano:2009yt, Beneke:2010dz, Anisimov:2010dk, Garbrecht:2011aw, Garny:2011hg, Drewes:2012ma, Garbrecht:2012pq, Frossard:2012pc, Iso:2013lba,Garny:2010nj, Drewes:2013gca, Garbrecht:2013iga, Hohenegger:2013zia,
Hohenegger:2014cpa,Garbrecht:2014aga, Iso:2014afa, Anisimov:2008dz, Anisimov:2010aq, Beneke:2010wd, Garbrecht:2010sz}.  In particular, these `first-principles' approaches to leptogenesis can, in principle,   account consistently for all off-shell, finite-width and flavour effects, including thermal corrections. However, the
loopwise perturbative  expansion of non-equilibrium  propagators 
is      normally     spoiled      by  mathematical  pathologies, known as pinch
singularities \cite{Weldon:1991ek, Altherr:1994fx, Altherr:1994jc, Bedaque:1994di, Dadic:1998yd, Greiner:1998ri, Garbrecht:2011xw},  arising from  ill-defined products  of Dirac delta
functions  with identical  arguments. Thus, in order to define and extract physically-meaningful quantities, such as particle
number densities, one often resorts to particular approximations, specifically gradient expansion of time-derivatives in the so-called Wigner representation \cite{Winter:1986da, Berges:2005md, Garbrecht:2008cb, Garny:2010nz, Vlasenko:2013fja} and quasi-particle ansaetze \cite{Lipavsky:1986zz, Bornath:1996zz, Herranen:2010mh, Herranen:2011zg, Fidler:2011yq} for the form of the dressed propagators.

Recently,  a new  perturbative formulation of  non-equilibrium   thermal   field theory~\cite{Millington:2012pf} (for an overview, see~\cite{Millington:2013isa}) was developed, which is free of the mathematically ill-defined pinch singularities previously thought to spoil such approaches. Within this framework, one may define a perturbative loopwise truncation scheme for quantum transport equations that is valid to all orders in a gradient expansion, whilst capturing non-Markovian dynamics, memory and threshold effects. As a result, physically-meaningful particle number densities can be derived directly from the Noether charge, without the need for quasi-particle ansaetze.

In this paper, we use the formalism of~\cite{Millington:2012pf} to obtain a well-defined loopwise perturbative truncation scheme for the KB equations in the weakly-resonant regime of RL. This is achieved without resorting to a quasi-particle \ansatz~for the dressed heavy-neutrino propagator. We find that the source term for the lepton asymmetry obtained in this KB approach is exactly the same as that derived in~\cite{Dev:2014laa} using a semi-classical Boltzmann approach. Thus, we prove that there is {\em no} double-counting of flavour effects captured in the fully flavour-covariant semi-classical formalism of~\cite{Dev:2014laa}. Moreover, we confirm that flavour mixing and oscillations are two \emph{physically-distinct} phenomena also in improved quantum Boltzmann treatments, as is the case in the semi-classical approach. Finally, we show that the use of KB or quasi-particle ansaetze for the resummed heavy-neutrino propagators captures {\em only} flavour oscillations and {\em not} the separate phenomenon of flavour mixing. As a result, the application of such approximations may lead to an underestimate of the generated lepton asymmetry by a factor of order two in the weakly-resonant regime.

The rest of the paper is organized as follows. In Section~\ref{sec:model}, we introduce a flavour-covariant scalar toy model of RL and describe its flavour-covariant canonical quantization within the context of perturbative non-equilibrium thermal field theory. In Section~\ref{sec:source}, we derive the heavy-neutrino and charged-lepton quantum transport equations relevant to the source term for the lepton asymmetry. Subsequently, in Section~\ref{sec:KBansatz}, we illustrate that the KB ansaetze are not appropriate in the presence of particle mixing and that such approximations, when applied to the resummed heavy-neutrino propagator, discard the phenomenon of flavour mixing. We then proceed to derive the form of the source term for the lepton asymmetry, incorporating both flavour mixing and oscillation. In Section~\ref{sec:approx}, we derive an approximate analytic solution for the lepton asymmetry, making comparison with existing results. Finally, Section~\ref{sec:conc} summarizes our conclusions. Further technical details of the resummation of the heavy-neutrino Yukawa couplings and thermal propagators are provided in~\ref{app:resprop}.

\section{Flavour-Covariant Scalar Model of RL}\label{sec:model}

In order to study the role of heavy-neutrino flavour effects in RL within the KB formalism, but without the technical complications arising from the fermionic nature of the heavy neutrinos and charged leptons, we consider a simple toy model of RL with two real scalar fields $N_\alpha$ (with $\alpha = 1,2$),  one complex scalar field $L$ and a real scalar $\Phi$. This simple model includes all qualitatively important features of leptogenesis, where the two real scalar fields mimic heavy Majorana neutrinos of two flavours and the complex scalar field models charged leptons of a single flavour. Moreover, $\Phi$ plays the role of the the Standard Model (SM) Higgs field. The approximate global $U(1)$ symmetry associated with the complex scalar field $L$ corresponds to the lepton number. Similar toy models have been used extensively to study RL in the KB formalism \cite{Garny:2009rv,Garny:2009qn,Garny:2010nj,Garny:2010nz,Hohenegger:2013zia,Hohenegger:2014cpa}.

In order to capture fully the flavour-dynamics in the heavy-neutrino sector, we adopt the flavour-covariant formulation developed in \cite{Dev:2014laa}. Therein, the heavy-neutrino field transforms in the fundamental representation of $U(\mathcal{N}_{N})$, i.e.~$N_{\alpha} \to N'_{\alpha} = U_{\alpha}^{\phantom{\alpha}\beta}N_{\beta}$, where $U_{\alpha}^{\phantom{\alpha}\beta} \in U(\mathcal{N}_{N})$.\footnote{In this covariant notation, complex conjugation raises/lowers indices.} The relevant part of the Lagrangian may then be written in the following manifestly-covariant form:
\begin{equation}
\label{Lag}
\mathcal{L}_{N}\ =\ h^{\alpha}L^\dag \Phi N_{\alpha}\:+\:\frac{1}{4}N_{\alpha}[m_N^2]^{\alpha\beta}N_{\beta}\:+\:\mathrm{H.c.}\;,
\end{equation}
where the tree-level Yukawa coupling parameters $h^{\alpha}$ transform as a vector and the heavy-neutrino mass-squared matrix $[m^2_{N}]^{\alpha\beta}$ transforms as a rank-2 tensor of $U(\mathcal{N}_N)$, i.e.
\begin{equation}
\h{}{\alpha} \ \rightarrow \ h_{}'^{\alpha} \ = \ U^\alpha_{\phantom{\alpha} \beta} \, \h{}{\beta}\;, \qquad [m_N^2]^{\alpha \beta} \ \rightarrow \
[m'^{2}_N]^{\alpha \beta} \ = \ U^\alpha_{\phantom{\alpha} \gamma} \;
  U^\beta_{\phantom{\beta} \delta} \, [m^2_N]^{\gamma \delta} \;.
\end{equation}
The basic Sakharov conditions~\cite{Sakharov:1967dj} for the generation of the BAU are satisfied in this toy model as follows. The lepton number is explicitly broken by the $L^\dag \Phi N$ term in \eqref{Lag}. Also, the charge conjugation ($C$) symmetry is violated, provided that ${\rm arg}(h^1)\neq {\rm arg}(h^2)$ and the heavy neutrinos are non-degenerate. In this model, $C$-violation will also imply $\CP$-violation, since $\CP$-transformations on the scalar fields are identical to $C$-transformations, up to a sign change of the spatial coordinates. Finally, the out-of-equilibrium condition can be satisfied by the decays of $N_\alpha$ in an expanding Universe.   

\begin{figure}[t!]
\centering
\includegraphics[scale=1]{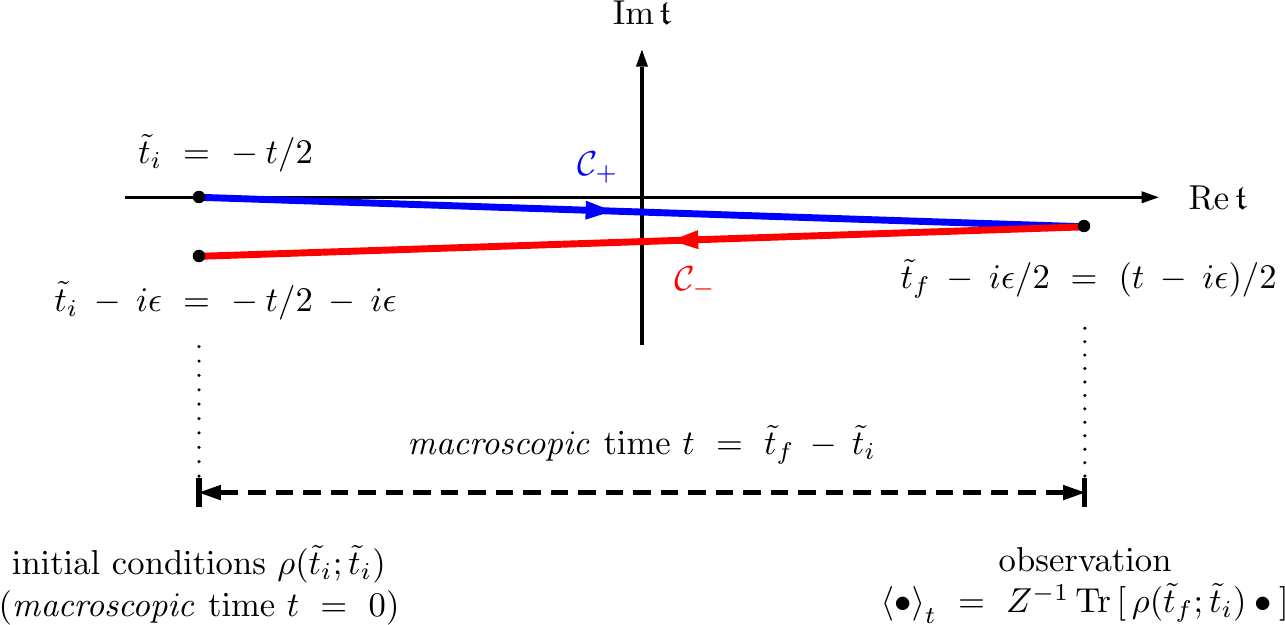}
\caption{The CTP contour $\mathcal{C}_+\cup\mathcal{C}_-$ in the analytically-continued complex-time ($\mathfrak{t}$) plane, indicating the relationship between macroscopic and microscopic times $t=\tilde{t}_f-\tilde{t}_i$, as defined in the perturbative non-equilibrium formulation of thermal Quantum Field Theory in \cite{Millington:2012pf}.\label{fig:contour}}
\end{figure}

In order to define unambiguously the physical objects that enter into the heavy-neutrino rate equation, we follow the perturbative framework of non-equilibrium thermal field theory described in \cite{Millington:2012pf}. This novel approach differs from the standard interpretation of the Schwinger-Keldysh closed-time path (CTP) formalism in the time-dependence of free propagators. Specifically, the free positive-frequency Wightman propagator of \cite{Millington:2012pf}, in the \emph{interaction picture}, is defined as
\begin{equation}
[i\Delta^{N, \, 0}_>(x,y,\tilde{t}_f;\tilde{t}_i)]_\alpha^{\phantom \alpha \beta} \ =\ \braket{N_{\alpha}(x;\tilde{t}_i)N^{\beta}(y;\tilde{t}_i)}_{t}\ \equiv\ \frac{1}{Z} \, \mathrm{Tr}\,\rho(\tilde{t}_f;\tilde{t}_i) N_\alpha(x;\tilde{t}_i)N^\beta(y;\tilde{t}_i)\;,
\label{prop1}
\end{equation}
where $Z=\mathrm{Tr}\,\rho(\tilde{t}_f;\tilde{t}_i)$ is the partition function, with $\rho(\tilde{t}_f;\tilde{t}_i)$ being the quantum-statistical density operator. Here, $\tilde{t}_f$ is the microscopic time of observation of the system and $\tilde{t}_i$ is the boundary time at which the three equivalent pictures of quantum mechanics, viz.~Schr\"{o}dinger, interaction (Dirac) and Heisenberg, are coincident and the initial conditions may be specified unambiguously. Together, these two microscopic times determine the macroscopic time of the statistical evolution: $t=\tilde{t}_f-\tilde{t}_i$. As a result, the statistical part of the free propagators evolves in time. Picture-independent physical observables may then be defined by taking the equal-time limit of ensemble expectation values. The resulting path-integral description is constructed over a modification of the original CTP contour, with $\tilde{t}_f=-\,\tilde{t}_i=t/2$, whose length evolves in time. This is illustrated graphically in Figure~\ref{fig:contour}.

This is in stark contrast with earlier CTP constructions (see e.g.~\cite{Calzetta:1986ey,Calzetta:1986cq}), which use the Heisenberg picture and a contour of fixed length. In these earlier treatments, the free propagator is given by 
\begin{equation}
[i\Delta^{N,\, 0}_{>}(x,y,0)]_\alpha^{\phantom \alpha \beta} \ =\ \braket{N_{\alpha}(x)N^{\beta}(y)}_{0}\ \equiv\ \frac{1}{Z} \, \mathrm{Tr}\,\rho(0) N_\alpha(x)N^\beta(y) \;,
\label{prop2}
\end{equation}
whose role is to encode the initial conditions at a time $t=0$.

With the recognition of the necessary dependence of diagrammatic series on the two microscopic times $\tilde{t}_f$ and $\tilde{t}_i$, it was shown in \cite{Millington:2012pf} that one may arrive at a perturbative framework of non-equilibrium field theory, using~\eqref{prop1}, that captures fully non-Markovian effects and is free of the so-called pinch singularities~\cite{Weldon:1991ek, Altherr:1994fx, Altherr:1994jc, Bedaque:1994di, Dadic:1998yd, Greiner:1998ri, Garbrecht:2011xw}, previously thought to spoil such perturbative approaches when constructed using~\eqref{prop2}. As a result, it is now possible to obtain a well-defined perturbative loopwise truncation scheme for quantum transport equations, using the propagator~\eqref{prop1} instead of~\eqref{prop2}. Moreover, it was illustrated that this loopwise perturbative truncation was two-fold, proceeding (i) {\em spectrally:} the truncation of the external leg of the transport equation determines the order of spectral dressing of the species being counted and (ii) {\em statistically:} the truncation of the self-energies determines the set of processes driving the statistical evolution of the system. In this way, quantum transport equations are obtained without the need for quasi-particle approximations or gradient expansion.

Within the union of these flavour-covariant and perturbative non-equilibrium frameworks \cite{Dev:2014laa,Millington:2012pf}, the plane-wave decomposition of the heavy-neutrino field takes the following form:
\begin{equation}
\label{Nfield}
N_{\alpha}(x;\tilde{t}_i)\ =\ \int_{\ve k} \Esdu{N}{\mathbf{k}}{\alpha}{\beta}\Big(\edu{-}{k}{\beta}{\gamma}a_{\gamma}(\mathbf{k},0;\tilde{t}_i)\:+\:\edu{+}{k}{\beta}{\gamma}G_{\gamma\delta}\,a^{\delta}(\mathbf{k},0;\tilde{t}_i)\Big)\;,
\end{equation}
where $\int_{\ve k}\equiv \int \!\! \frac{\mathrm{d}^3 \ve k}{(2\pi)^3}$ is a short-hand notation for the three-momentum integral. Here, the energy and Fourier kernels transform as rank-2 tensors under $U(\mathcal{N}_N)$, since $[E_N^2(\mathbf{k})]_{\alpha}^{\phantom{\alpha}\beta} = \mathbf{k}^2\delta_{\alpha}^{\phantom{\alpha}\beta}+[|m_N|^2]_{\alpha}^{\phantom{\alpha}\beta}$.\footnote{Here, $[|m_N|^2]_{\alpha}^{\phantom{\alpha}\beta} = [m_N]_{\alpha \gamma}[m_N]^{\gamma \beta}$. In the mass eigenbasis, $[m_N]^{\alpha \beta}$ is diagonal and its elements are the heavy-neutrino mass eigenvalues (for more details, see \cite{Dev:2014laa}).} In addition, we have been careful to indicate that interaction-picture annihilation and creation operators $a_{\alpha}(\mathbf{k},\tilde{t};\tilde{t}_i)$ and $a^{\alpha}(\mathbf{k},\tilde{t};\tilde{t}_i)$ depend explicitly on the time~$\tilde{t}$ and implicitly on the boundary time~$\tilde{t}_i$. The algebra of the creation and annihilation operators for the scalar fields is defined by the equal-time commutator
\begin{equation}
\label{com}
\big[a_{\alpha}(\mathbf{k},\tilde{t};\tilde{t}_i),\ a^{\beta}(\mathbf{k}',\tilde{t};\tilde{t}_i)\big]\ =\ (2\pi)^3\delta_{\alpha}^{\phantom{\alpha}\beta}\delta^{(3)}(\mathbf{k}-\mathbf{k}')\;,
\end{equation}
where $\delta_{\alpha}^{\phantom{\alpha}\beta}$ is the Kronecker delta. Notice that we have chosen the normalization of the operators (having mass dimensions $-3/2$) such that the commutator \eqref{com} is isotropic in flavour space. The unitary and symmetric matrix $\mat{G}$, with elements $G_{\alpha \beta} = [U^* U^\dag]_{\alpha \beta}$,   appearing in \eqref{Nfield}, is required by flavour covariance, since the operators $a^{\alpha}(\mathbf{k},\tilde{t};\tilde{t}_i)$ and $a_{\alpha}(\mathbf{k},\tilde{t};\tilde{t}_i)$ necessarily transform in different representations of $U(\mathcal{N}_N)$. Moreover, charge-conjugate pairs of creation or annihilation operators must also transform in different representations of $U(\mathcal{N}_N)$ (for a detailed discussion, see \cite{Dev:2014laa}). This requires us to introduce the generalized charge-conjugation ($\gC$) transformation, defined via
\begin{equation}
[a_{\alpha}(\mathbf{k},\tilde{t};\tilde{t}_i)]^{\gC}\ \equiv\ G^{\alpha\beta}[a_{\beta}(\mathbf{k},\tilde{t};\tilde{t}_i)]^C\ = \ G^{\alpha\beta}\,\mathcal{U}_C\,a_{\beta}(\mathbf{k},\tilde{t};\tilde{t}_i)\,\mathcal{U}_C^{\dag}\ =\ G^{\alpha\beta}a_{\beta}(\mathbf{k},\tilde{t};\tilde{t}_i)\;,
\end{equation}
where $\mathcal{U}_C$ is the charge-conjugation operator in Fock space. Thus, $\mat{G}$ accounts for flavour rotations via $U$ to and from the mass eigenbasis, in which the usual charge-conjugation $C$ is defined. In the mass eigenbasis, which we denote by a caret ($\widehat{\phantom n}$), we have $\widehat{\mat{G}} = \bm{1}_{2}$, in which case the $\gC$ and $C$ transformations coincide. We may now write the generalized ``Majorana'' constraint
\begin{equation}
\label{Majoranacon}
[N^{\alpha}]^{\gC}\ =\ N_{\alpha}\;.
\end{equation}
In addition, for this toy model, $\gC \equiv C$ for the charged-lepton and Higgs fields. Finally, we note that the heavy-neutrino Yukawa couplings transform as
\begin{equation}
(\h{}{\alpha})^{\gC} \ = \ \hs{}{\alpha}\;,
\end{equation}
ensuring that the Lagrangian in \eqref{Lag} has definite $\gC$ properties.

We now introduce the heavy-neutrino number densities
\begin{subequations}
\begin{gather}
[n^N(\mathbf{k},t)]_{\alpha}^{\phantom{\alpha}\beta}\ =\ \mathcal{V}_3^{-1}\,\braket{a^{\beta}(\mathbf{k},\tilde{t};\tilde{t}_i)a_{\alpha}(\mathbf{k},\tilde{t};\tilde{t}_i)}_t\;,\\
[\overline{n}^N(\mathbf{k},t)]_{\alpha}^{\phantom{\alpha}\beta}\ =\ \mathcal{V}_3^{-1}\,G_{\alpha\mu}\braket{a^{\mu}(\mathbf{k},\tilde{t};\tilde{t}_i)a_{\lambda}(\mathbf{k},\tilde{t};\tilde{t}_i)}_tG^{\lambda\beta}\ =\ [(n^N(\mathbf{k},t))^{\gC}]_{\beta}^{\phantom \beta \alpha}\;,
\end{gather}
\end{subequations}
where $\mathcal{V}_3\equiv (2\pi)^3\delta^{(3)}(\mathbf{p}=\ve{0})$ is the spatial 3-volume. Notice that $n^N$ and $\overline{n}^N$ are not independent by virtue of the (real scalar) ``Majorana'' constraint \eqref{Majoranacon}. We may then define the $\gC$ ``even'' and ``odd'' heavy-neutrino number densities~\footnote{We adopt the notation of~\cite{Dev:2014laa}, where the bold-face $\mat{A}$ denotes the entire matrix in flavour space, while $[A]_{\alpha}^{\phantom{\alpha}\beta}$ denotes its individual elements.} 
\begin{equation}
\underline{\mat{n}}^N\ \equiv \ \frac{1}{2}\big(\mat{n}^N + \overline{\mat{n}}^N\big)\;,\qquad \mat{\delta n}^N\ =\ \mat{n}^N - \overline{\mat{n}}^N\;.
\end{equation}
In the mass eigenbasis, these become
\begin{equation}
\underline{\widehat{\mat{n}}}^N\ \ = \ \Re\big[\widehat{\mat{n}}^N \big] \;, \qquad \mat{\delta} \widehat{\mat{n}}^N \ = \ 2i \, \Im\big[\widehat{\mat{n}}^N \big] \;.
\end{equation}
Lastly, we introduce the flavour-covariant generalized real and imaginary parts, which, for an Hermitian matrix $\mat{A} = \mat{A}^\dag$, are defined as
\begin{equation}
[\widetilde{\Re}(A)]_\alpha^{\phantom \alpha \beta} \ \equiv \ \frac{1}{2} \big(A_\alpha^{\phantom \alpha \beta} \, +\, G_{\alpha \mu} \, A_{\lambda}^{\phantom \lambda \mu} \, G^{\lambda \beta} \big) \;,\qquad [\widetilde{\Im}(A)]_\alpha^{\phantom \alpha \beta} \ \equiv \ \frac{1}{2i} \big(A_\alpha^{\phantom \alpha \beta} \, -\, G_{\alpha \mu} \, A_{\lambda}^{\phantom \lambda \mu} \, G^{\lambda \beta} \big) \;.
\end{equation}

In the weakly-resonant regime of RL, i.e.~for $\Gamma_{N_{1,2}}\ll |m_{N_1}-m_{N_2}|\ll m_{N_{1,2}}$, the heavy-neutrino mass eigenbasis can be defined to be that in which the thermal mass matrix, given in terms of the retarded self-energy $i\mat{\Pi}_{\mathrm{R}}^N(k)$ by 
\begin{equation}\label{eq:thermal_mass}
[M^2_N(k)]_{\alpha}^{\phantom \alpha \beta} \ \equiv \ [|m_N|^2]_{\alpha}^{\phantom \alpha \beta} \;-\; [\widetilde{\Re} \, \Pi_{\mathrm{R}}^N(k) ]_{\alpha}^{\phantom \alpha \beta}\;,
\end{equation}
is diagonal in the vicinity of the two quasi-degenerate thermal mass shells. This is based on the fact that the equilibrium retarded heavy-neutrino self-energy $i\mat{\Pi}_{\mathrm{R},\,\mathrm{eq}}^N(k)$ is a slowly-varying function of $k_0$ near the thermal mass shells, so that the mass eigenbasis is well defined. Therefore, we can approximate $[M^2_N(k)]_{\alpha}^{\phantom \alpha \beta}$ by its on-shell (OS) form $[M^2_N(\ve k)]_{\alpha}^{\phantom \alpha \beta}$, given by the solution of the thermal gap equation at equilibrium
\begin{equation}
\label{gap}
[\mathscr{E}^2_N(\ve k)]_{\alpha}^{\phantom{\alpha}\beta}\ \equiv\ {\ve k}^2\delta_{\alpha}^{\phantom{\alpha}\beta}+[M^2_N(\ve k)]_{\alpha}^{\phantom \alpha \beta} \ = \ [E_N^2(\mathbf{k})]_{\alpha}^{\phantom{\alpha}\beta} \;-\; \lim_{\epsilon \to 0^+} [\widetilde{\Re} \, \Pi_{\mathrm{R},\,\mathrm{eq}}^N(\mathscr{E}_N(\ve k)+i\epsilon,\ve k)]_{\alpha}^{\phantom \alpha \beta}  \;.
\end{equation}

Assuming a Gaussian and spatially-homogeneous ensemble for the heavy neutrinos, we may write the double-momentum representation (see \cite{Millington:2012pf,Dev:2014laa}) of the heavy-neutrino Wightman propagators in the mass eigenbasis as
\begin{align}
\label{Wfull}
&[i\widehat{\Delta}^{N,\,0}_{\gtrless}(k,k',\tilde{t}_f;\tilde{t}_i)]_{\alpha\beta}\ =\ 2\pi|2k_0|^{1/2}\delta(k^2-\widehat{m}_{N,\,\alpha}) \, 2\pi|2k_0'|^{1/2}\delta(k'^2-\widehat{m}_{N,\,\beta}) \, e^{i(k_0-k_0')\tilde{t}_f}\nonumber\\& ~
\times 
  \Big(\theta(\pm k_0)\theta(\pm k_0')\delta_{\alpha\beta} + [\theta(k_0)\theta(k_0')+\theta(-k_0)\theta(-k_0')][\widehat{n}^N(\mathbf{k},t)]_{\alpha\beta}\Big)(2\pi)^3\delta^{(3)}(\mathbf{k}-\mathbf{k}').
\end{align}
Here, we see that, in general, the heavy-neutrino Wightman propagators depend explicitly on the zeroth components of two four momenta, $k_0$ and $k_0'$, since the time-translational invariance of free propagators is broken in the presence of flavour coherences. The phase $e^{i(k_0-k_0')\tilde{t}_f}$ arises from the free evolution of the interaction-picture operators.

In the weakly-resonant regime, we may approximate $\widehat{m}_{N,\,\alpha} \simeq m_N=(\widehat{m}_{N,\,1}+\widehat{m}_{N,\,2})/2$ in the on-shell delta functions of \eqref{Wfull}.  We then obtain the free homogeneous heavy-neutrino Wightman propagators, which, in a general basis, may be written in the single momentum representation
\begin{equation}
\label{homogenN}
[i \Delta^{N,\,0}_{\gtrless}(k,t)]_{\alpha}^{\phantom \alpha \beta}\ =\ 2\pi\delta(k^2-m_{N}^2)
  \Big(\theta(\pm k_0)\delta_{\alpha}^{\phantom{\alpha}\beta}\:+\:[n^N(\mathbf{k},t)]_{\alpha}^{\phantom \alpha \beta}\Big)\;.
\end{equation}
By resumming the dispersive self-energy corrections, we may replace $m_N$ in \eqref{homogenN} by the average thermal mass $M_N(\ve k)\equiv (\widehat{M}_{N,\,1}(\ve k)+\widehat{M}_{N,\,2}(\ve k))/2$, given by the solution to \eqref{gap}.

For our subsequent discussion, we need in addition the equilibrium form of the dressed Higgs and charged-lepton Wightman propagators for  vanishing chemical potential. In the narrow-width approximation (NWA), it will be sufficient to use the standard quasi-particle expressions
\begin{align}
\label{LPhiprops1}
i\Delta^{\Phi, \, \mathrm{eq}}_{\gtrless}(q)\ &=\ 2\pi\delta(q^2-M_{\Phi}^2) \, \big[\theta(\pm q_0)\:+\:n^{\Phi}_{\mathrm{eq}}(\mathbf{q}))\big]\;,\\
\label{LPhiprops2}
i\Delta^{L, \, \mathrm{eq}}_{\gtrless}(p)\ &=\ 2\pi\delta(p^2-M_L^2) \, \big[\theta(\pm p_0)\:+\:\theta(p_0)n^{L}_{\mathrm{eq}}(\mathbf{p})\:+\:\theta(-p_0)\overline{n}^{L}_{\mathrm{eq}}(\mathbf{p})\big]\;,
\end{align}
where $M^2_X$ denotes the thermal mass of the species $X$ and $n^X_{\rm eq}(\ve p) = (e^{\mathscr{E}_X(\ve p)/T} - 1)^{-1}$ is the equilibrium number density of $X$, obeying Bose-Einstein statistics, with $\mathscr{E}_X(\ve p)$ being the OS quasi-particle energy, as determined by a thermal gap equation analogous to~\eqref{gap}.

\section{Quantum Transport Equations}\label{sec:source}

In this section, we will obtain the rate equations for the heavy neutrinos and charged leptons, derived within the perturbative framework of \cite{Millington:2012pf}, as outlined above. In particular, we will derive the form of the source term for the charged-lepton asymmetry in the scalar toy model described in Section~\ref{sec:model}.

Employing the methods described in \cite{Millington:2012pf}, we may define the total number density unambiguously in terms of the negative-frequency Wightman propagator as
\begin{equation}
\label{numdef}
\bm{n}(t,\mathbf{X})\ = \ \int^{(X)}_{p,\,p'}(p_0+p_0')\,i\bm{\Delta}_<(p,p',\tilde{t}_f;\tilde{t}_i)\;,
\end{equation}
where we have introduced the following short-hand notation for the integration measure:
\begin{equation}
\label{measure}
\int^{(X)}_{p,\,p'}\ \equiv\ \int_{p,\,p'}e^{-i(p-p')\cdot X}\,\theta(p_0+p_0') \;,
\end{equation}
with $\int_p \equiv \int \!\!\frac{\D{4}{p}}{(2 \pi)^4}$ and $\int_{p,p', \,\ldots} \equiv \int_p \int_{p'} \cdots$. Here, $X\equiv X^\mu= (\tilde{t}_f,\mathbf{X})$ is the macroscopic space-time coordinate four-vector. Notice that the definition \eqref{numdef} is valid to any order in a perturbative truncation of the heavy-neutrino propagator. By inserting the free heavy-neutrino propagator on the RHS of \eqref{numdef}, we obtain the number density $n^N$ of spectrally-free particles (with respect to absorptive transitions). Instead, inserting the resummed heavy-neutrino propagator, we count the number density $n^N_{\mathrm{dress}}$ of fully spectrally-dressed particles.

In coordinate space, the KB equations for the Wightman propagators of a given species may be written in the following condensed form (see e.g.~\cite{Prokopec:2003pj}):
\begin{align}
\label{KB1}
\Big(-\Box_x^2\:-\: |\bm{m}|^2\cdot\:+\: \bm{\Pi}_{\mathcal{P}}\ast\Big)\bm{\Delta}_{\gtrless} \ = \ -\:\frac{1}{2}\,\Big(\bm{\Pi}_>\ast\bm{\Delta}_<\:-\:\bm{\Pi}_<\ast\bm{\Delta}_>\:+\:2\,\bm{\Pi}_{\gtrless}\ast\bm{\Delta}_{\mathcal{P}}\Big) \;,
\end{align}
where $\Box_x^2\equiv\partial_{x^{\mu}}\partial_{x_{\mu}}$ is the d'Alembertian operator, $\ast$ indicates the convolution
\begin{equation}
\label{ast}
\bm{A}\ast\bm{B}\ \equiv\ \int_{z\,\in\,\Omega_t}\bm{A}(x,z,\tilde{t}_f;\tilde{t}_i)\cdot \bm{B}(z,y,\tilde{t}_f;\tilde{t}_i)\;,
\end{equation}
and $\cdot$ denotes matrix multiplication in flavour space. Here, $\int_{z\,\in\,\Omega_t} \equiv \int_{\Omega_t}d^4z$ is the space-time convolution integral over the hypervolume $\Omega_t=[\tilde{t}_i,\tilde{t}_f]\times\mathbb{R}^3=[-\frac{t}{2},\frac{t}{2}]\times\mathbb{R}^3$, bounded temporally by the boundary and observation times \cite{Millington:2012pf}. In addition, we note that
\begin{equation}
\bm{B}\ast\bm{A}\ \equiv\ \int_{z\,\in\,\Omega_t}\bm{B}(x,z,\tilde{t}_f;\tilde{t}_i)\cdot \bm{A}(z,y,\tilde{t}_f;\tilde{t}_i)\;,
\end{equation}
{\em without} reversal of the external arguments $x$ and $y$. In~\eqref{KB1}, $i\bm{\Pi}_{>(<)}$ are the absorptive self-energies arising from unitarity cuts with positive- (negative-) energy flow, whilst $i\bm{\Pi}_{\mathcal{P}}$ and $i\bm{\Delta}_{\mathcal{P}}$ are the principal-part self-energy and propagator, respectively. For the propagators and self-energies of the charged-lepton and Higgs, the matrix product ($\cdot$) trivially reduces to scalar multiplication. 

Performing a double Fourier transform (see~\cite{Millington:2012pf}), \eqref{KB1} takes the following double-momentum representation:
\begin{align}
\label{KB2}
\Big(p^2\:-\: |\bm{m}|^2\cdot\:+\: \bm{\Pi}_{\mathcal{P}}\star\Big)\bm{\Delta}_{\gtrless} \ = \ -\:\frac{1}{2}\,\Big(\bm{\Pi}_>\star\bm{\Delta}_<\:-\:\bm{\Pi}_<\star\bm{\Delta}_>\:+\:2\,\bm{\Pi}_{\gtrless}\star\bm{\Delta}_{\mathcal{P}}\Big) \;,
\end{align}
where $\star$ denotes the weighted convolution integral in the double momentum space
\begin{equation}
\label{star}
\bm{A}\star\bm{B}\ \equiv\ \int_{q,\,q'}\; (2\pi)^4\delta^{(4)}_t(q-q')\,\bm{A}(p,q,\tilde{t}_f;\tilde{t}_i)\cdot \bm{B}(q',p',\tilde{t}_f;\tilde{t}_i)\;.
\end{equation}
Here, the weight function is given by
\begin{equation}
(2\pi)^4\delta^{(4)}_t(q-q')\ \equiv\ \int_{z\,\in\,\Omega_t}e^{-i(q-q')\cdot z}\ =\ (2\pi)^4\delta_t(q-q')\delta^{(3)}(\mathbf{q}-\mathbf{q}')\;,
\end{equation}
with
\begin{equation}
\delta_t(q_0-q_0')\ \equiv\ \frac{1}{\pi}\frac{\sin[(q_0-q_0')t/2]}{q_0-q_0'}\;.
\end{equation}
As for the $\ast$ operation in \eqref{ast}, the external arguments $p$ and $p'$ are not reversed for $\bm{B}\star\bm{A}$ relative to \eqref{star}.

Following \cite{Millington:2012pf} and using \eqref{numdef}, we find the rate equation for the total number density
\begin{align}
\label{KB3}
&\frac{\D{}{\bm{n}(t,\mathbf{X})}}{\D{}{t}}\:-\:\int_{p,\,p'}^{(X)}(\mathbf{p}^2-\mathbf{p}'^2)\,\bm{\Delta}_{<}\:-\: \int_{p,\,p'}^{(X)}\Big([|\bm{m}|^2,\ \bm{\Delta}_{<}]\:-\:[\bm{\Pi}_{\mathcal{P}},\ \bm{\Delta}_{<}]_{\star}\Big) \nonumber\\&\qquad =\ -\:\frac{1}{2}\int_{p,\,p'}^{(X)}\Big(\{\bm{\Pi}_{>},\ \bm{\Delta}_{<}\}_{\star}\:-\:\{\bm{\Pi}_{<},\ \bm{\Delta}_{>}\}_{\star}\:+\:2\,[\bm{\Pi}_{<},\ \bm{\Delta}_{\mathcal{P}}]_{\star}\Big)\;.
\end{align}
Here, we have introduced the (anti-)commutators in flavour space:
\begin{subequations}
\begin{align}
[\bm{A},\, \bm{B}]_{\star}\ &\equiv\ \bm{A}\star\bm{B}\:-\:\bm{B}\star\bm{A}\;,\\
\{\bm{A},\, \bm{B}\}_{\star}\ &\equiv\ \bm{A}\star\bm{B}\:+\:\bm{B}\star\bm{A}\;,
\end{align}
\end{subequations}
with the $\star$ operation defined in \eqref{star} above. In \eqref{KB3}, the first two terms of the LHS comprise the drift terms; the latter two terms of the LHS describe mean-field effects, including oscillations; and, finally, the terms on the RHS describe collisions. We emphasize that \eqref{KB3} is obtained without the need to perform a gradient expansion or make use of a quasi-particle ansatz. Thus,~\eqref{KB3} is valid at {\em any} order in perturbation theory for spatially inhomogeneous systems, thereby capturing {\em fully} the flavour effects, non-Markovian dynamics and memory effects.  

\subsection{Heavy-Neutrino Rate Equations}
\label{Nrates}

Starting from the general transport equation \eqref{KB3}, we now proceed to derive the rate equation for the heavy-neutrino number densities. The principal-part self-energy $\bm{\Pi}_{\mathcal{P}}^N$ in the last term on the LHS of \eqref{KB3} combines with the tree-level heavy-neutrino mass $|\bm{m}_N|^2$ to give the thermal mass: $\mat{M}_N^2 = |\mat{m}_N|^2 - \mat{\Pi}_\mathcal{P}^N$, where we have used $\widetilde{\Re}(\mat{\Pi}_{\mathrm{R}}^N) =  \mat{\Pi}_\mathcal{P}^N$ in~\eqref{eq:thermal_mass}. In the absence of mixing, the commutator containing $\mat{\Delta}^N_\mathcal{P}$ involves a principal-value integral that we may safely neglect for quasi-degenerate heavy neutrinos. Nevertheless, mixing between the Majorana neutrinos causes the appearance of off-diagonal entries in $\mat{\Delta}^N_\mathcal{P}$ proportional to Dirac delta functions in the NWA. It can be shown that, in the weakly-resonant regime, these are higher-order effects compared to the ones taken into account in our analysis. 

Following \cite{Millington:2012pf} and using the definition of the total number density in \eqref{numdef}, we obtain the following rate equation for the dressed heavy-neutrino number density $n^N_{\mathrm{dress}}$:
\begin{align}\label{eq:dressed_n}
&\frac{\D{}{}{\bm{n}}^{N}_{\mathrm{dress}}}{\D{}{t}} \ = \ \int^{(X)}_{k,\,k'}\: \bigg[ - \,i\, \big[\bm{M}^2_N,\ i \bm{\Delta}_<^N\big]\: - \:  \frac{1}{2}\Big( \big\{i \bm{\Pi}^N_<,\ i \bm{\Delta}^N_>\big\}_{\star}\:-\: \big\{i \bm{\Pi}^N_>,\ i \bm{\Delta}_<^N\big\}_{\star}\Big) \Bigg] \;.
\end{align}
Neglecting the $O(h^6)$ terms proportional to the lepton asymmetry, we may approximate the charged-lepton and Higgs propagators in the heavy-neutrino self-energies by their quasi-particle equilibrium forms, as given in \eqref{LPhiprops1} and \eqref{LPhiprops2}. The non-Markovian heavy-neutrino self-energies may then be written in the form
\begin{align}
& [i\Pi_{\gtrless}^N(k,k',\tilde{t}_f;\tilde{t}_i)]_{\alpha}^{\phantom{\alpha}\beta} \  =\ 2\,\widetilde{\Re}\,(h^\dag h)_{\alpha}^{\phantom \alpha \beta}\nonumber \\& \qquad \qquad \times \ \int_{p,\,q}(2\pi)^4\delta_t(k-p-q)\,(2\pi)^4\delta_t(k'-p-q)\,\Delta^{L,\mathrm{eq}}_\lessgtr(p) \, \Delta^{\Phi,\mathrm{eq}}_\lessgtr(q)\;.
\end{align}

We now perform a Wigner-Weisskopf approximation along the lines of \cite{Dev:2014laa}, in order to obtain the Markovian limit of \eqref{eq:dressed_n}. The Wigner-Weisskopf approximation is performed by the replacement of $\Omega_t$ by $\Omega_{\infty}$ in the space-time integrals, which corresponds to taking the limit $t\to\infty$ in the vertex functions by virtue of the identity
\begin{equation}
\lim_{t\to\infty}\delta_t(k_0-p_0-q_0)\ =\ \delta(k_0-p_0-q_0)\;.
\label{appx1}
\end{equation}
We note that the free-phase contributions in \eqref{measure} and those present in the dressed heavy-neutrino propagator, cf.~\eqref{Wfull}, will cancel in this energy-conserving limit. Thus, we make the following replacement of the dressed heavy-neutrino propagator in the Markovian approximation
\begin{equation}
e^{-i(k_0-k_0')\tilde{t}_f}\bm{\Delta}^N_{<}(k,k',\tilde{t}_f;\tilde{t}_i) \ \longrightarrow \ \bm{\Delta}^N_{<}(k,k',t)\;,
\label{appx2}
\end{equation}
where the latter is distinguished by the form of its time argument.

With the approximations \eqref{appx1} and \eqref{appx2}, we obtain from \eqref{eq:dressed_n} the Markovian heavy-neutrino rate equation for the dressed number density:
\begin{align}\label{eq:KB_N_2}
\frac{\D{}{}[{n}^{N}_{\rm dress}]_{\alpha}^{\phantom{\alpha}\beta}}{\D{}{t}} \ &= \ \int_{k,\,k'} \theta(k_0+k_0')\: \bigg[ - \,i\, \big[M^2_N,  \, i \Delta_<^{N}(k,k',t)\big]_{\alpha}^{\phantom \alpha \beta} \notag\\ & \quad - \frac{1}{2} \, \Big(  [i \Pi^N_<(k)]_{\alpha}^{\phantom \alpha \gamma} \; [i \Delta^{N}_>(k,k',t)]_{\gamma}^{\phantom \gamma \beta} \; + \; [i \Delta^{N}_>(k,k',t)]_{\alpha}^{\phantom \alpha \gamma} \;  [i \Pi^N_<(k')]_{\gamma}^{\phantom \gamma \beta} \Big) \notag\\
& \quad + \frac{1}{2} \, \Big(  [i \Pi^N_>(k)]_{\alpha}^{\phantom \alpha \gamma} \; [i \Delta^{N}_<(k,k',t)]_{\gamma}^{\phantom \gamma \beta} \; + \; [i \Delta^{N}_<(k,k',t)]_{\alpha}^{\phantom \alpha \gamma} \;  [i \Pi^N_>(k')]_{\gamma}^{\phantom \gamma \beta} \Big) \bigg] \;,
\end{align}
in which the explicit forms of the Markovian heavy-neutrino self-energies are given by
\begin{align}
i [\Pi^N_{\lessgtr}(k)]_{\alpha}^{\phantom \alpha \beta} \ &= \ 2\, \widetilde{\Re}(h^\dag h)_{\alpha}^{\phantom \alpha \beta} \, B_\lessgtr^{\mathrm{eq}}(k)\;.
\end{align}
Herein, we have introduced the thermal loop functions
\begin{equation}
B^{\mathrm{eq}}_{\lessgtr}(k) \ \equiv \ \int_{p,\,q} \,(2 \pi)^4 \, \delta^{(4)}(p-k+q) \, \Delta_{\lessgtr}^{\Phi,\mathrm{eq}}(q) \, \Delta^{L,\mathrm{eq}}_{\lessgtr}(p) \;,
\end{equation}
which satisfy $B^{\mathrm{eq}}_{<}(-k) = B^{\mathrm{eq}}_{>}(k) \in \mathbb{R}$. In the classical-statistical regime and restricting to positive energy flow ($k_0>0$), the thermal loop functions may be written as
\begin{align}
B^{\mathrm{eq}}_{>}(k_0>0,\ve k) \ & = \ - \, \int \D{}{\Pi_\Phi} \int \D{}{\Pi_L} \, (2 \pi)^4 \, \delta^{(4)}(k-p_\Phi-p_L) \;, \label{eq:th_loop1}\\
B^{\mathrm{eq}}_{<}(k_0>0, \ve k) \ & = \ - \, \int \D{}{\Pi_\Phi} \int \D{}{\Pi_L} \, (2 \pi)^4 \, \delta^{(4)}(k-p_\Phi-p_L) \, n^\Phi_{\rm eq}(E_\Phi) \, n^L_{\rm eq}(E_L)\;. \label{eq:th_loop2}
\end{align}
The phase space measure appearing here is defined as
\begin{equation}
\D{}{\Pi}_X \ \equiv \ \frac{\D{4}{p_X}}{(2 \pi)^4} \, 2\pi \delta(p_X^2 - M_X^2) \, \theta(p_X^0) \;,
\end{equation}
for a given species $X$. Notice that \eqref{eq:KB_N_2} still accounts for the non-homogeneity of the heavy-neutrino propagator.

Following~\cite{Millington:2012pf} and as described in Section~\ref{sec:model}, the rate equations \eqref{KB3} may be truncated in a perturbative loopwise manner as follows: (i) spectrally, by truncating the external leg of the KB equation and (ii) statistically, by truncating the self-energy. In order to obtain the asymmetry at $O(h^4)$, it is sufficient to consider the evolution of the spectrally-free heavy-neutrino number density. This is obtained by truncating \eqref{eq:KB_N_2} spectrally at zeroth loop order, replacing the external heavy-neutrino propagators by the free homogeneous propagator in \eqref{homogenN}. On the other hand, we retain the full 1-loop CJT-resummed statistical evolution, described by the heavy-neutrino self-energy, which contains the dressed (quasi-particle) charged-lepton and Higgs propagators in \eqref{LPhiprops1} and \eqref{LPhiprops2}. We then obtain the rate equation for the spectrally-free number density (denoted as $[n^N]_{\alpha}^{\phantom{\alpha}\beta}$)
\begin{align}\label{eq:KB_N_3}
&\frac{\D{}{}[{n}^{N}]_{\alpha}^{\phantom{\alpha}\beta}}{\D{}{t}} \ = \ \int_k \theta(k_0) \bigg\{ - \,i\, \big[M^2_N,  \, i \Delta_<^{N,\,0}(k,t)\big]_{\alpha}^{\phantom \alpha \beta} \nonumber\\&\qquad \qquad \qquad \qquad  - \;  \frac{1}{2} \, \Big( \big\{i \Pi^N_<(k),\, i \Delta^{N,\,0}_>(k,t)\big\}_{\alpha}^{\phantom \alpha \beta}\,-\, \big\{i \Pi^N_>(k),\, i \Delta_<^{N,\,0}(k,t)\big\}_{\alpha}^{\phantom \alpha \beta}\Big) \bigg\} \;,
\end{align}
where the $k'$-integral in \eqref{eq:KB_N_2} was carried out trivially. 

Substituting explicitly for the form of the free heavy-neutrino propagator in \eqref{homogenN} and assuming kinetic equilibrium along the lines of \cite{Dev:2014laa}, \eqref{eq:KB_N_3} gives the rate equation for $n^N(t)$:
\begin{equation}\label{eq:evol_N}
  \frac{\D{}{}[{n}^{N}]_{\alpha}^{\phantom{\alpha}\beta}}{\D{}{t}}
  \ = \ - \; i \,
  \Big[\mathcal{E}_N,\, n^{N}\Big]_\alpha^{\phantom \alpha \beta} \;
  + \; \Tdu{\big[\widetilde{\rm Re}
    (\gamma^{N,(0)}_{L \Phi})\big]}{}{}{\alpha}{\beta} \;
  - \; \frac{1}{2 \, n^N_{\rm eq}} \,
  \Big\{{n}^N, \, \widetilde{\rm Re}(\gamma^{N,(0)}_{L \Phi})
  \Big\}_{\alpha}^{\phantom{\alpha}\beta} \; ,
\end{equation}
where the $\gCP$-``even'' rate is defined in terms of the tree-level Yukawa couplings
\begin{equation}
[\gamma^{N,(0)}_{L \Phi}]_{\alpha}^{\phantom{\alpha}\beta} \ \equiv \ \int_{N L \Phi} 2 \, h_\alpha h^\beta \;,
\end{equation}
with the short-hand notation
\begin{equation}
\int_{NL\Phi} \ \equiv \ \int \D{}{\Pi_N}  \int \D{}{\Pi_L} \int \D{}{\Pi_\Phi} \, (2 \pi)^4 \, \delta^{(4)}(p_N-p_L-p_\Phi) \, e^{-p_N^0/T} \;.
\end{equation}
The thermally-averaged effective energy matrix is \cite{Dev:2014laa}
\begin{equation}
\mat{\mathcal{E}}_N \ = \ \frac{g_N}{n^N_{\rm eq}} \, \int_{\ve k} \, \pmb{\mathscr{E}}_{\!N}(\ve k) \, e^{-\mathscr{E}_N(\ve k)/T} \; ,
\end{equation}
with $\pmb{\mathscr{E}}_{\!N}(\ve k)$ defined in~\eqref{gap}. Here, $\mathscr{E}_N(\ve k)=(|\ve k|^2+M_N^2)^{1/2}$ is the average energy and $M_N=\frac{1}{2}{\rm Tr}(\mat{M}_N^\dag \mat{M}_N)$ is the average thermal mass for the system of two quasi-degenerate heavy neutrinos. Moreover, we have indicated explicitly the dependence on the number of internal degrees of freedom of the heavy neutrino scalars $g_N=1$, in order to facilitate the comparison with the realistic case of Majorana fermions, where $g_N=2$.
Separating the $\gCP$-``even'' and -``odd'' parts of \eqref{eq:evol_N}, we obtain the final rate equations
\begin{subequations}
  \label{nfin}
\begin{align}
  \frac{\D{}{}[\underline{n}^{N}]_{\alpha}^{\phantom{\alpha}\beta}}{\D{}{t}}
  \ & = \ - \; \frac{i}{2} \,
  \Big[\mathcal{E}_N,\, \delta n^{N}\Big]_\alpha^{\phantom \alpha \beta} \;
  + \; \Tdu{\big[\widetilde{\rm Re}
    (\gamma^{N,(0)}_{L \Phi})\big]}{}{}{\alpha}{\beta} \;
  - \; \frac{1}{2 \, n^N_{\rm eq}} \,
  \Big\{\underline{n}^N, \, \widetilde{\rm Re}(\gamma^{N,(0)}_{L \Phi})
  \Big\}_{\alpha}^{\phantom{\alpha}\beta}
   \;, \label{eq:evol_n}\\[3pt]
  \frac{\D{}{[\delta n^N]_\alpha^{\phantom \alpha \beta}}}{\D{}{t}} \ &
  = \ - \; 2 \, i \, 
  \Big[\mathcal{E}_N,\, \underline{n}^{N}\Big]
  _\alpha^{\phantom \alpha \beta} \; - \; \frac{1}{2 \, n^N_{\rm eq}}  \,
  \Big\{\delta n^N, \, \widetilde{\rm Re}(\gamma^{N,(0)}_{L \Phi})
  \Big\}_{\alpha}^{\phantom{\alpha}\beta}\;, \label{eq:evol_dn} 
\end{align}
\end{subequations}
which agree with those obtained in the semi-classical approach of~\cite{Dev:2014laa}, when the effective Yukawa couplings used there are replaced by the tree-level ones. As we will show below, the flavour-covariant rate equations in \eqref{nfin} are sufficient to obtain the form of the lepton asymmetry at $O(h^4)$ in the weakly-resonant regime, in complete agreement with the results presented in \cite{Dev:2014laa}. In particular, we draw attention to the second term on the RHS of \eqref{eq:evol_n}, as identified in \cite{Dev:2014laa}, which induces flavour coherences in the heavy-neutrino number density $[\underline{n}^N]_{\alpha}^{\phantom{\alpha}\beta}$, thus triggering oscillations \emph{in addition} to mixing.

\subsection{Lepton Asymmetry Source Term}

The lepton asymmetry is defined in terms of the total number densities of the charged leptons and anti-leptons, $n^L$ and $\overline{n}^{L}$, as
\begin{equation}
\delta n^L\ \equiv\ n^L \:-\:\overline{n}^L\;.
\end{equation}
The source term for this asymmetry is obtained by considering the contribution to the lepton transport equation that contains the $\CP$-even part of the (anti-)lepton and Higgs propagators. In the regime where the asymmetry is small, we may approximate these propagators as having the equilibrium forms given by \eqref{LPhiprops1} and \eqref{LPhiprops2} in the single-momentum representation.

Proceeding analogously to the heavy-neutrino case, replacing the charged-lepton and Higgs propagators by their quasi-particle equilibrium forms in \eqref{LPhiprops1} and \eqref{LPhiprops2}, we obtain the following Markovian approximation of the source term for the lepton asymmetry:\footnote{For further details of the diagrammatic representation of non-homogeneous self-energies and, in particular, their double-momentum structure, see \cite{Millington:2012pf,Millington:2013isa}.}
\begin{align}
\label{eq:KB_Mark}
\frac{\D{}{\delta n^L} }{\D{}{t}} \  &\supset \ -\:i \int_{k,k',\,p,\,q} \theta(p_0+k_0'-q_0)(2\pi)^4\delta^{(4)}(p-k+q)\nonumber\\&\qquad\qquad \times\Big[ \hs{}{\beta} \h{}{\alpha} \Big(\Del{<}{k,k',t}{\alpha}{\beta} \, \, \Delta_{>}^{\Phi,\,\mathrm{eq}}(q) \, \Delta^{L,\,\mathrm{eq}}_>(k'-q)\nonumber\\&\qquad \qquad \qquad \qquad -\: \Del{>}{k,k',t}{\alpha}{\beta} \, \Delta_{<}^{\Phi,\,\mathrm{eq}}(q) \, \Delta^{L,\,\mathrm{eq}}_<(k'-q) \Big)\: -\: \gC\!.c. \Big] \;,
\end{align}
where $\gC\!.c.$ denotes the generalized charge-conjugate terms. 

In the next section, we will demonstrate explicitly that it is not appropriate to replace the non-homogeneous heavy-neutrino propagator in \eqref{eq:KB_Mark} by the homogeneous approximation of the {\em free} heavy-neutrino propagator given in \eqref{homogenN}. We note that this would correspond to a statistical truncation of the source term for the lepton asymmetry $\delta n^L$ and \emph{not} a spectral truncation, as was the case with this replacement in the heavy-neutrino rate equations of Section~\ref{Nrates} [cf. \eqref{eq:KB_N_3}].

\section{Flavour Mixing and Kadanoff--Baym Ansaetze}
\label{sec:KBansatz}

In this section, we will derive the contribution of the dressed heavy-neutrino Wightman propagators to the source term for the asymmetry in the presence of flavour mixing. In addition, we will show that the standard quasi-particle or KB ansaetze for the form of these propagators are insufficient to capture all physically-relevant phenomena. Specifically, we will demonstrate how both heavy-neutrino mixing and oscillations provide {\em distinct} contributions to the $O(h^4)$ lepton asymmetry in the weakly-resonant regime and that the flavour mixing contribution is tacitly discarded when the standard quasi-particle or KB ansaetze are used.

\begin{figure}[p]
\centering
\vspace{-3em}
\begin{align*}
\raisebox{-1em}{\includegraphics[scale=0.6]{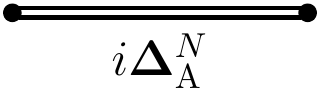}}\ &=\ \raisebox{-1.1em}{\includegraphics[scale=0.6]{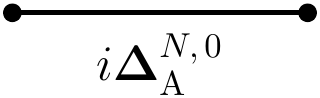}}\:+\:\raisebox{-1.1em}{\includegraphics[scale=0.6]{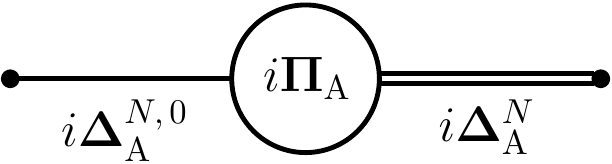}}\\[24pt]
\raisebox{-1em}{\includegraphics[scale=0.6]{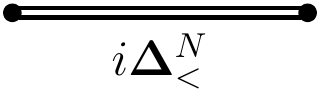}}\ &=\ \raisebox{-1.1em}{\includegraphics[scale=0.6]{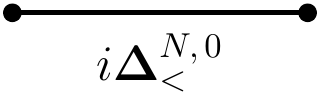}}\nonumber\\
&\qquad +\raisebox{-1.1em}{\includegraphics[scale=0.6]{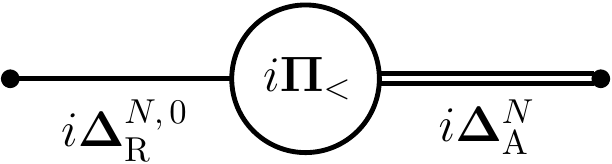}}+ \raisebox{-1.1em}{\includegraphics[scale=0.6]{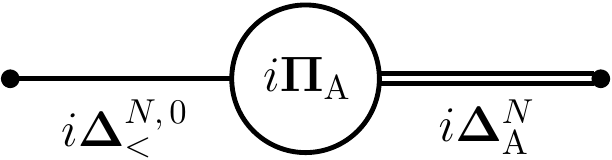}}+\raisebox{-1.1em}{\includegraphics[scale=0.6]{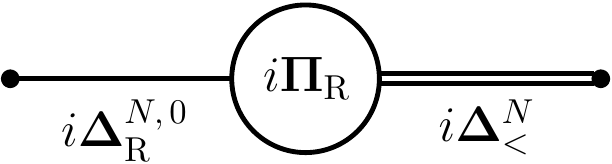}}\\[12pt]
& = \ \raisebox{-1.1em}{\includegraphics[scale=0.6]{less0}}\nonumber\\[3pt]
&\qquad+\raisebox{-1.1em}{\includegraphics[scale=0.6]{lessblob1}}+\raisebox{-1.1em}{\includegraphics[scale=0.6]{lessblob2}}+\raisebox{-1.1em}{\includegraphics[scale=0.6]{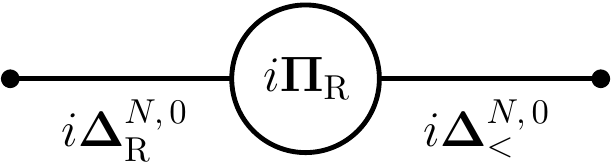}}\\[3pt]
&\qquad +\raisebox{-1.1em}{\includegraphics[scale=0.6]{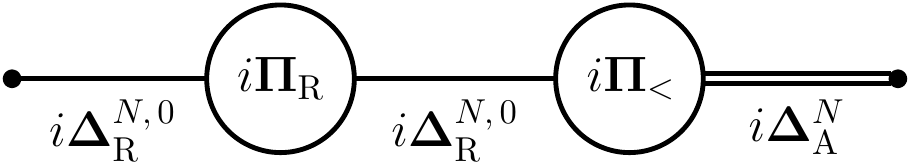}}+\raisebox{-1.1em}{\includegraphics[scale=0.6]{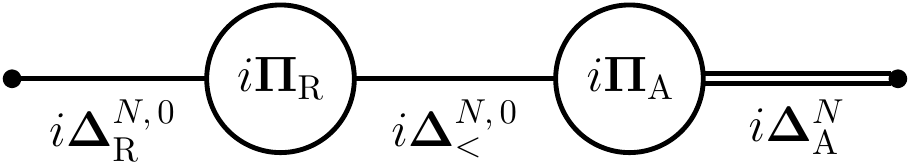}}\\[3pt]
&\qquad +\raisebox{-1.1em}{\includegraphics[scale=0.6]{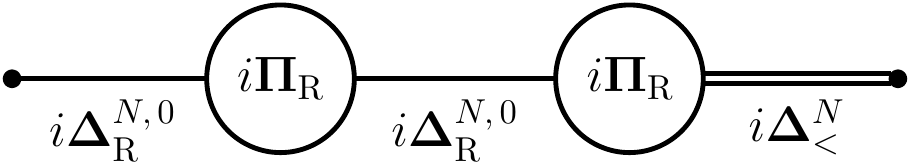}}\\[12pt]
& = \ \bigg(\raisebox{0em}{\includegraphics[scale=0.6]{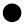}}+\raisebox{-1.1em}{\includegraphics[scale=0.6]{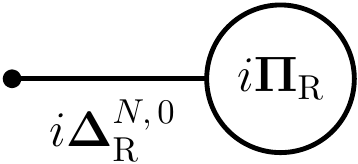}}+\raisebox{-1.1em}{\includegraphics[scale=0.6]{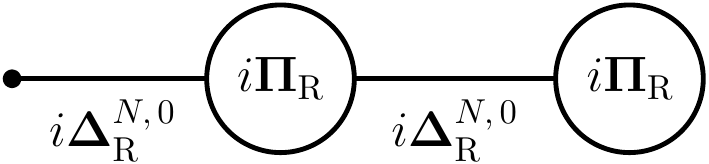}}+\ldots\bigg) \;\times \; \raisebox{-1em}{\includegraphics[scale=0.6]{less0}} \\[3pt]
& \qquad \qquad \times \; \bigg(\ldots+\raisebox{-1.1em}{\includegraphics[scale=0.6]{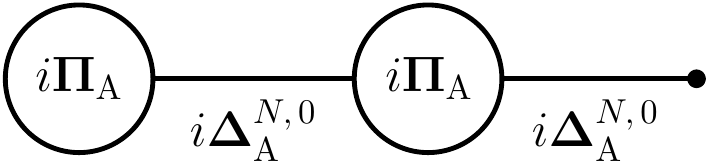}}+\raisebox{-1.1em}{\includegraphics[scale=0.6]{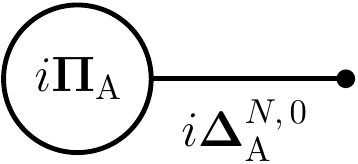}}+\raisebox{0em}{\includegraphics[scale=0.6]{dot}}\bigg)\\[6pt]
& \qquad + \; \bigg(\raisebox{-1.1em}{\includegraphics[scale=0.6]{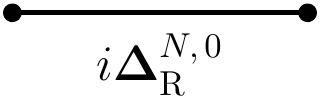}}+\raisebox{-1.1em}{\includegraphics[scale=0.6]{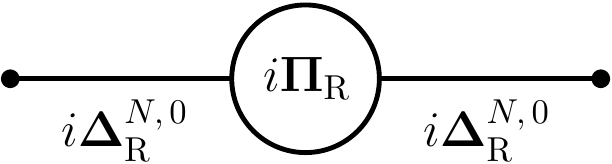}}+\cdots\bigg) \; \times \; \raisebox{-1.1em}{\includegraphics[scale=0.6]{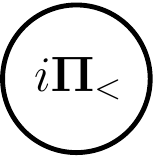}}\\[3pt]
&\qquad \qquad \times \; \bigg(\raisebox{-1.1em}{\includegraphics[scale=0.6]{adv0}} + \raisebox{-1.1em}{\includegraphics[scale=0.6]{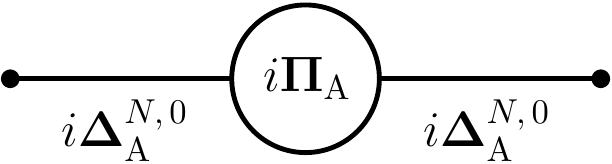}} + \cdots\bigg)\\[12pt]
& \equiv \quad \raisebox{-1.1em}{\includegraphics[scale=0.6]{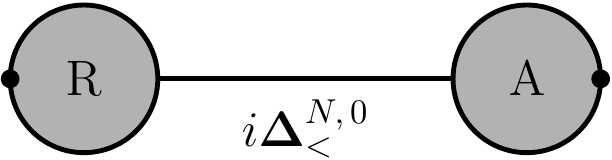}} \quad + \quad \raisebox{-1.1em}{\includegraphics[scale=0.6]{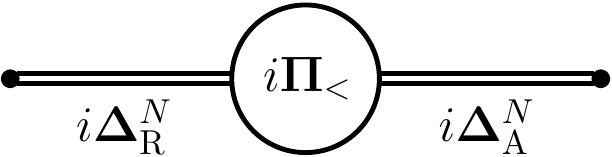}}
\end{align*}
\caption{Diagrammatic representation of the iterative solution to the Schwinger-Dyson equation for the dressed heavy-neutrino matrix Wightman propagator $i\bm{\Delta}_<^{N}$. Here, the double lines are fully dressed propagators, whereas the single lines are the  propagators dressed with dispersive corrections only. The unshaded circles are the relevant self-energies, whereas the shaded ones are the amputated self-energy corrections to the vertices, which can be identified at leading order with the resummed Yukawa couplings (see~\eqref{eq:diag} and~\ref{app:resprop}). \label{fig:iterations}}
\end{figure}

In the Markovian approximation and assuming that the charged-lepton and Higgs propagators have the equilibrium forms in \eqref{LPhiprops1} and \eqref{LPhiprops2}, the Schwinger-Dyson equation of the dressed heavy-neutrino Wightman propagator takes the form \cite{Millington:2012pf}
\begin{align}\label{eq:SD}
&i \mat{\Delta}^{N}_{<}(k,k',t) \ = \ i \mat{\Delta}^{N,\, 0}_{<}(k,k',t)\:+\: i \mat{\Delta}^{N,\, 0}_{\mathrm{R}}(k) \cdot i \mat{\Pi}_<(k)(2\pi)^4\delta^{(4)}(k-k') \cdot  i \mat{\Delta}^{N}_{\mathrm{A}}(k')\nonumber\\&\qquad+\: i \mat{\Delta}^{N,\, 0}_{\mathrm{R}}(k) \cdot i \mat{\Pi}_{\mathrm{R}}(k) \cdot  i \mat{\Delta}^{N}_{<}(k,k',t) \:+\: i \mat{\Delta}^{N,\, 0}_{<}(k,k',t) \cdot i \mat{\Pi}_{\mathrm{A}}(k') \cdot  i \mat{\Delta}^{N}_{\mathrm{A}}(k') \;.
\end{align}
Instead, the equation for the advanced propagator takes on the simple closed form:
\begin{equation}
i \mat{\Delta}^{N}_{\mathrm{A}}(k) \ = \ i \mat{\Delta}^{N,\, 0}_{\mathrm{A}}(k) \:+\: i \mat{\Delta}^{N,\, 0}_{\mathrm{A}}(k) \cdot i \mat{\Pi}_{\mathrm{A}}(k) \cdot  i \mat{\Delta}^{N}_{\mathrm{A}}(k) \;.
\end{equation}
As shown diagrammatically in Figure~\ref{fig:iterations}, we can solve \eqref{eq:SD} iteratively, obtaining
\begin{align}
\label{fullWight}
&[i\Delta^{N}_{<}(k,k',t)]_{\alpha}^{\phantom{\alpha}\beta}\ =\  [i\Delta^N_{\mathrm{R}}(k)]_{\alpha}^{\phantom{\alpha}\gamma}[i\Pi^N_<(k)]_{\gamma}^{\phantom{\gamma}\delta}(2\pi)^4\delta^{(4)}(k-k')[i\Delta^N_{\mathrm{A}}(k')]_{\delta}^{\phantom{\delta}\beta}\nonumber\\&\quad +\:\sum_{m\:=\:0}^{\infty}\Big[\big(i\Delta^0_{\mathrm{R}}(k)\cdot i\Pi^N_{\mathrm{R}}(k)\big)^{m}\Big]_{\alpha}^{\phantom{\alpha}\gamma} \, [i\Delta_<^{N,\,0}(k,k',t)]_{\gamma}^{\phantom{\gamma}\delta}\sum_{n\:=\:0}^{\infty}\Big[\big(i\Pi^N_{\mathrm{A}}(k')\cdot i\Delta^{N,\,0}_{\mathrm{A}}(k') \big)^n\Big]_{\delta}^{\phantom{\delta}\beta}\;.
\end{align}
Note that \eqref{fullWight} is free of pinch singularities, since we have been accounting for the violation of time-translational invariance (for a more detailed discussion, see \cite{Millington:2012pf}). An alternative derivation of the homogeneous Markovian form of the dressed Wightman propagator is given in \ref{app:resprop} by means of a direct matrix inversion, which is in agreement with \eqref{fullWight}. We also discuss the NWA of these dressed CTP propagators in \ref{app:resprop}.

The first term on the RHS of \eqref{fullWight} gives the washout due to $\Delta L = 0$ and $\Delta L = 2$ scatterings. Notice that it does not contribute to the source term, since, if we extract the latter by taking the equilibrium part of $\mat{\Pi}_<$, the whole term has an equilibrium form at $O(h^4)$, as considered here. Therefore, \emph{no double-counting is present} and an explicit real intermediate state (RIS) subtraction~\cite{Kolb:1980} procedure is not needed, as expected on general grounds in the KB formalism~\cite{Garny:2009rv, Garny:2009qn, Beneke:2010wd}. The second term on the RHS of \eqref{fullWight} can be written in terms of the resummed Yukawa couplings $\mathbf{h}^{\alpha}$ \cite{Pilaftsis:2003gt, Dev:2014laa}, since, at leading order, we have the following equivalence in the heavy-neutrino mass eigenbasis:
\begin{align}
&\widehat{h}^\alpha \bigg[ \sum_{n\:=\:0}^\infty \Big ( i \widehat{\Delta}_{\mathrm{R}}^0(k) \cdot i \widehat{\Pi}_{\mathrm{R}}(k) \Big)^n \bigg]_{\alpha}^{\phantom \alpha \beta} \ = \ \raisebox{-1.3em}{\includegraphics[scale=0.45]{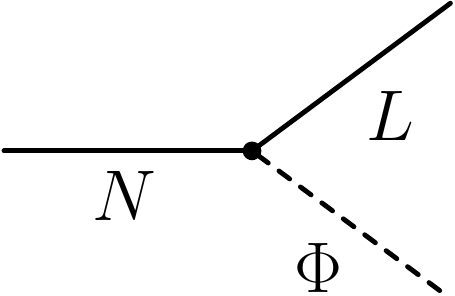}}\ +\ \raisebox{-1.3em}{\includegraphics[scale=0.45]{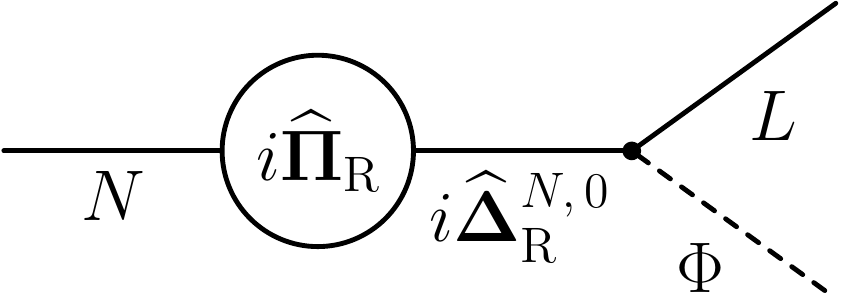}}\nonumber\\[1em]&\hspace{8em} +\ \raisebox{-1.3em}{\includegraphics[scale=0.45]{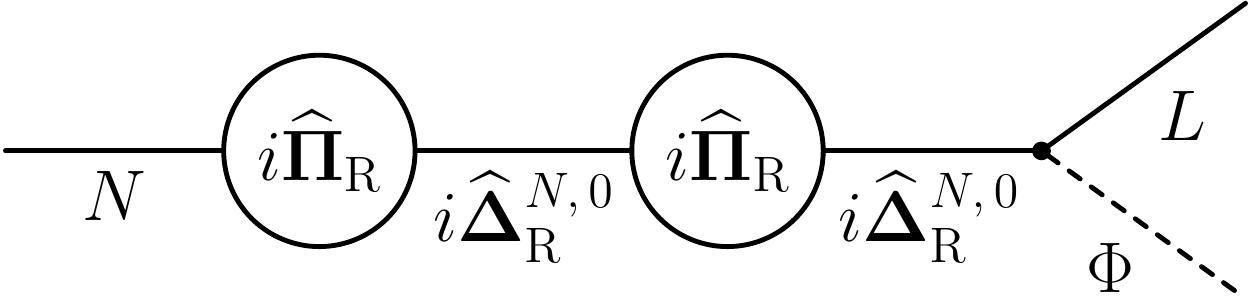}}\ +\ \cdots \quad \ \sim \ \widehat{\mathbf{h}}^\beta \;.
\label{eq:diag}
\end{align}
In \ref{app:res_Yuk}, we prove that this equivalence holds in \eqref{eq:KB_Mark} at $O(h^4)$ in the lepton asymmetry, at least in the weakly-resonant regime. There, we also show that, in the mass eigenbasis, the part of the heavy-neutrino propagator contributing to the source term for the asymmetry is
\begin{align}
\label{fullWight2}
&[i\widehat{\Delta}^{N}_{<}(k,k',t)]_{\alpha \beta}\ \supset  \ [\widehat{\Delta}_{\mathrm{R}}^N(k)]_{\alpha \gamma} \, \Big( [\widehat{\Delta}_{\mathrm{R}}^{N,\,0}(k)]^{-1}_{\gamma \gamma}  \, [i \widehat{\Delta}_<^{N,\,0}(k,k',t)]_{\gamma \delta} \, [\widehat{\Delta}_{\mathrm{A}}^{N,\,0}(k')]^{-1}_{\delta \delta}\Big) \, [\widehat{\Delta}_{\mathrm{A}}^N(k')]_{\delta \beta}\;.
\end{align}

On the other hand, the KB ansatz for the heavy-neutrino propagator (restricting to positive frequencies) takes the following form in the heavy-neutrino mass eigenbasis:  
\begin{equation}
[i\widehat{\Delta}^{N}_{\mathrm{KB},\,<}(k,k',t)]_{\alpha\beta}\ =\ 2\pi\delta(k^2-\widehat{M}^2_{N,\,\alpha}) \, 2\pi\delta(k'^2-\widehat{M}^2_{N,\,\beta}) \, [n^N_{\mathrm{KB}}(\mathbf{k},t)]_{\alpha\beta} \, (2\pi)^3\delta^{(3)}(\mathbf{k}-\mathbf{k}')\;,
\end{equation}
which satisfies the following properties
\begin{align}
\label{KBzero}
\big(k^2-\widehat{M}^2_{N,\,\alpha}\big)\, [i\widehat{\Delta}^{N}_{\mathrm{KB},\,<}(k,k',t)]_{\alpha\beta}\ =\ 0\;,\qquad [i\widehat{\Delta}^{N}_{\mathrm{KB},\,<}(k,k',t)]_{\alpha\beta}\,\big(k'{}^2-\widehat{M}^2_{N,\,\beta}\big)\ =\ 0\;.
\end{align}
It is immediately apparent that the full form of the dressed heavy-neutrino Wightman propagator in \eqref{fullWight} and, equivalently, \eqref{fullWight2} does not satisfy \eqref{KBzero}, by virtue of the mixing that gives rise to the resummed Yukawa couplings. We therefore come to the following conclusion: \emph{the application of KB ansaetze for the heavy-neutrino propagators discards the physical phenomena of flavour mixing.}  

In \cite{Garbrecht:2011xw}, it has been pointed out that, for a single-flavour case, one needs to include explicitly the effect of the width of the heavy neutrinos in the collision terms, when performing a zeroth-order gradient expansion or, equivalently, the Markovian approximation. Our results, obtained in a different approach as compared to \cite{Garbrecht:2011xw}, show that the inclusion of off-diagonal widths in the source terms is also necessary in order to  describe properly the phenomena of flavour mixing.

Other approaches in the literature \cite{Garny:2011hg, Hohenegger:2014cpa}, although not relying explicitly on a KB ansatz, are still able to solve the KB equations for the dressed heavy-neutrino propagator only up to an unknown function that parametrizes the external perturbation of the system. Both mixing and oscillations are in principle present in such \emph{double-time} approaches. It is however not clear whether our predictions are in quantitative agreement, since a direct comparison is made difficult by the simplified non-equilibrium setting in a non-expanding Universe studied in \cite{Garny:2011hg, Hohenegger:2014cpa}.

\begin{figure}[t]
\begin{equation*}
\parbox{11em}{\includegraphics[width = 11em]{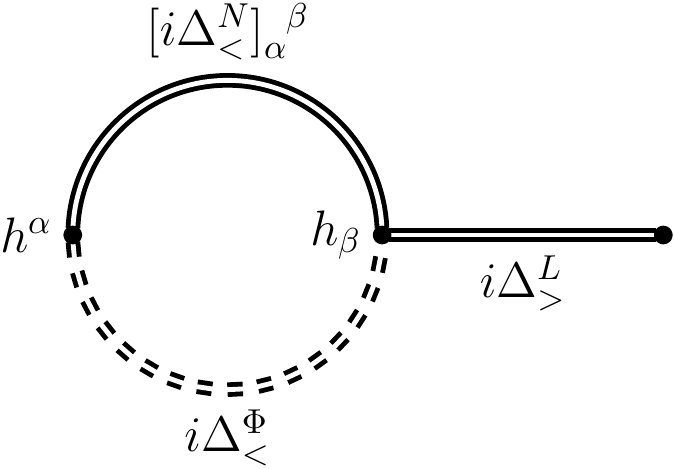}} \ \supset \ \parbox{11em}{\includegraphics[width = 11em]{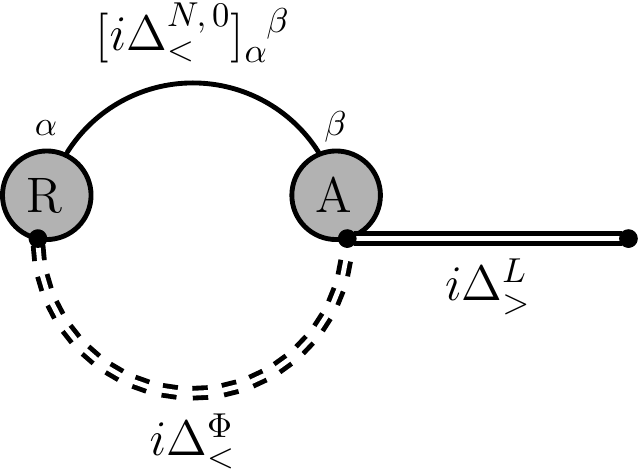}} \ \simeq \ \parbox{11em}{\includegraphics[width = 11em]{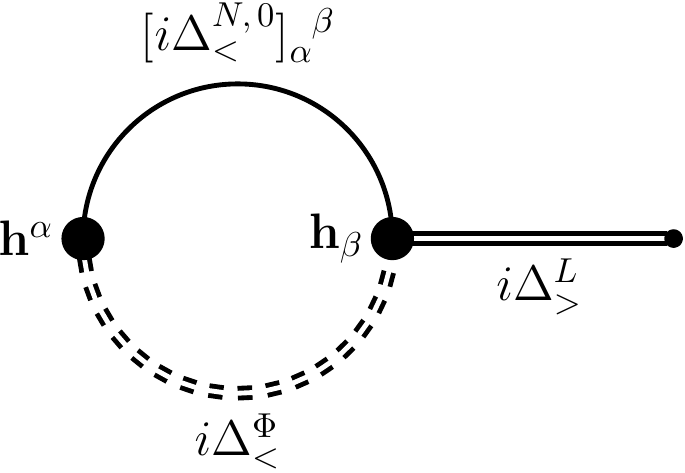}} 
\end{equation*}
\caption{Diagrammatic representation of the source term for the charged-lepton asymmetry in terms of the resummed Yukawa couplings and the spectrally-free heavy-neutrino propagator. \label{fig:self}}
\end{figure}

From~\eqref{eq:diag}, we see that the mixing effect due to the absorptive part of the heavy-neutrino self-energy in the dressed propagator~\eqref{fullWight} can be factorized into the resummed Yukawa couplings. Moreover, we can replace the non-homogeneous free heavy-neutrino propagator $\mat{\Delta}_<^{N,0}(k,k',t)$ on the RHS of~\eqref{fullWight} with the homogeneous approximation given by \eqref{homogenN}.  Thus, the contribution of the charged-lepton self-energy to the source term may be written in terms of the spectrally-free heavy-neutrino propagator and the resummed Yukawa couplings $\mathbf{h}^\alpha$ as  
\begin{align}
\label{eq:source_res_Yuk}
\frac{\D{}{\delta n^L} }{\D{}{t}} \  &\supset \ -\int_k \theta(k_0) \Big[ \hrs{}{\beta} \hr{}{\alpha} \Big([i\Delta^{N,\,0}_<(k,t)]_{\alpha}^{\phantom \alpha \beta} \, B_{>}^{\mathrm{eq}}(k)- [i\Delta^{N,\,0}_>(k,t)]_{\alpha}^{\phantom \alpha \beta} \, B_{<}^{\mathrm{eq}}(k) \Big)-\gC\!. c. \Big] \;,
\end{align}
without having required a quasi-particle ansatz for the dressed heavy-neutrino propagator. This procedure is illustrated diagrammatically in Figure~\ref{fig:self} and proven explicitly in \ref{app:res_Yuk}. 

Finally, assuming kinetic equilibrium\footnote{This is a valid assumption in the strong-washout regime of RL, since elastic scattering processes rapidly equilibrate the momentum distributions for all the relevant particle species on time-scales much shorter than those of their statistical evolution.} and separating the $\gCP$-``even'' and ``-odd'' parts of the heavy-neutrino number density $\mat{\underline{n}^{N}}$ and $\mat{\delta n^N}$, as described in \cite{Dev:2014laa}, the equation for the asymmetry becomes
\begin{equation}\label{eq:evol_dL}
\frac{\D{}{\delta n^L}}{\D{}{t}} \ 
  = \ 
  \bigg(\frac{[\underline{n}^{N}]_{\alpha}^{\phantom{\alpha}\beta}}
  {n^N_{\rm eq}} \, - \, \delta_{\alpha}^{\phantom{\alpha}\beta}\bigg) \,
  \Tdu{[\delta \gamma^{N}_{L \Phi}]}{}{}{\beta}{\alpha} \;
  + \; \frac{[\delta n^N]_{\alpha}^{\phantom{\alpha}\beta}}{2\,n^N_{\rm eq}} \,
  \Tdu{[\gamma^{N}_{L \Phi}]}{}{}{\beta}{\alpha}
  \;+\; \mathrm{W}[\delta n^L] \;,
\end{equation}
where $\mathrm{W}[\delta n^L]$ denotes the washout terms not studied explicitly here. The thermally-averaged rates are defined as
\begin{subequations}
\begin{align}
[\gamma^{N}_{L \Phi}]_{\alpha}^{\phantom{\alpha}\beta} \ & \equiv \ \int_{N L \Phi}  \big( \hrs{}{\alpha} \hr{}{\beta} \, +\, \hrc{}{\alpha} \hrcs{}{\beta}\big) \;, \\[0.5em]
[\delta \gamma^{N}_{L \Phi}]_{\alpha}^{\phantom{\alpha}\beta} \ &\equiv \ \int_{N L \Phi}  \big( \hrs{}{\alpha} \hr{}{\beta} \, -\, \hrc{}{\alpha} \hrcs{}{\beta}\big) \;,
\end{align}
\end{subequations}
where $\tilde{c}\equiv \widetilde{C}\!P$ indicates the generalized $\CP$ conjugate.
Equation \eqref{eq:evol_dL} describes the generation of the asymmetry via \emph{both} heavy-neutrino mixing (proportional to $[\delta \gamma^{N}_{L \Phi}]_{\alpha}^{\phantom{\alpha}\beta}$) and oscillations (proportional to $[\delta n^N]_{\alpha}^{\phantom{\alpha}\beta}$). In particular, the source terms agree with the ones given in \cite{Dev:2014laa}, where both the phenomena are separately identified and taken into account in the calculation of the final asymmetry.

\section{Approximate Analytic Solution for the Lepton Asymmetry}
\label{sec:approx}

In this section, we obtain an approximate analytic solution for the charged-lepton asymmetry in the strong-washout regime, using the KB rate equations derived in Section~\ref{sec:source}.
To this end, we introduce the following notational conventions:
\begin{equation}
\mat{\eta}^X \ \equiv \ \frac{\mat{n}^X}{n^\gamma}\;, \qquad \widehat{\mat{\upeta}}^N \ \equiv \ \frac{\widehat{\mat{\eta}}^N}
  {\eta^N_{\rm eq}} \, - \, \mat{1} \;, \qquad \widehat{\mathbf{K}} \ \equiv \ \frac{\widehat{\mat{\Gamma}}}{\zeta(3) H_N} \;,
\end{equation}
where $n^\gamma=2T^3\zeta(3)/\pi^2$ is the photon number density (with $\zeta(3)\approx 1.20206$) and $H_N$ is the Hubble constant in a Friedmann-Robertson-Walker Universe at temperature 
$T = M_N$. 
The thermal width $\mat{\Gamma}$ of the heavy neutrinos, obtained from $\widetilde{\Im}\: \mat{\Pi}^N_{\mathrm{R}} = M_N \mat{\Gamma}$, is related to the decay rate by means of $\widehat{\Pi}^N_<(k_0>0, \ve k) \simeq 2 i \, e^{-k_0/T} \Im\: \widehat{\Pi}^N_{\mathrm{R}}$, which implies
\begin{equation}
\widetilde{\Re} \, \mat{\gamma}^N_{L \Phi} \ = \ \frac{g_N}{2} \frac{M_N^3\, K_1(z) \, \mat{\Gamma}}{\pi^2 z} \;,
\label{5.2}
\end{equation}
with $z = M_N/T$ and $K_n(z)$ being the $n$-th order modified Bessel function of the second kind. We emphasize again that the off-diagonal elements of \eqref{5.2} induce flavour coherences in the heavy-neutrino sector via the second term on the RHS of \eqref{eq:evol_n}, giving rise to oscillations by virtue of the flavour commutators in \eqref{nfin}.

Taking into account the expansion of the Universe and using $\eta^N_{\rm eq} \approx g_N z^2 K_2(z)/4$, \eqref{eq:evol_n} and \eqref{eq:evol_dn} can be combined into \cite{Dev:2014laa}
\begin{equation}\label{eq:upeta_N}
  \frac{\D{}{\widehat{\mat{\upeta}}^N}}{\D{}{z}} \ 
  = \ \frac{K_1(z)}{K_2(z)} \, \bigg( \mat{1} \; 
  + \; \widehat{\mat{\upeta}}^N \; 
  - \; iz \, \bigg[\frac{\widehat{\mat{M}}_N}{\zeta(3)H_N}, \, 
  \widehat{\mat{\upeta}}^N \bigg] \, 
  - \, \frac{z}{2} \, \Big\{ \widehat{\mathbf{K}}, \, 
  \widehat{\mat{\upeta}}^N \Big\} \bigg) \;.
\end{equation}
In the strong-washout regime $[\mathrm{K}]_{\alpha \beta} \gg 1$, the system evolves towards the attractor solution given by
\begin{equation}\label{eq:upeta_N_attractor}
  i \, \bigg[\frac{\widehat{\mat{M}}_N}
  {\zeta(3)H_N}, \, \widehat{\mat{\upeta}}^N \bigg] \, 
  + \, \frac{1}{2} \, \Big\{\widehat{\mathbf{K}}^N, \,
  \widehat{\mat{\upeta}}^N \Big\} \ \simeq \ \frac{\mat{1}_2}{z} \;.
\end{equation}
The elements needed in what follows are found to be
\begin{align}
[\widehat{\upeta}^N]_{\alpha \alpha} \ &= \ \frac{1}{[\widehat{\mathrm{K}}^{(0)}]_{\alpha \alpha} \,z} \; \frac{(M_{N,\,1}-M_{N,\,2})^2 \, +\, (\widehat{\Gamma}^{(0)}_{11}+\widehat{\Gamma}^{(0)}_{22})^2/4}{(M_{N,\,1}-M_{N,\,2})^2 \, +\, \frac{(\widehat{\Gamma}^{(0)}_{11}+\widehat{\Gamma}^{(0)}_{22})^2 \, \Im[(\widehat{h}^\dag \widehat{h})_{12}]^2}{4 \, (\widehat{h}^\dag \widehat{h})_{11} (\widehat{h}^\dag \widehat{h})_{22}}} \ \simeq \ \frac{1}{[\widehat{\mathrm{K}}^{(0)}]_{\alpha \alpha} \,z} \;, \\[0.5em]
\Im[\widehat{\upeta}^N]_{12} \ &= \ \frac{\zeta(3) H_N}{z}\; \frac{[\widehat{\Gamma}^{(0)}]_{12}}{[\widehat{\Gamma}^{(0)}]_{11}[\widehat{\Gamma}^{(0)}]_{22}} \; \frac{(M_{N,\,1}-M_{N,\,2}) (\widehat{\Gamma}^{(0)}_{11}+\widehat{\Gamma}^{(0)}_{22})/2}{(M_{N,\,1}-M_{N,\,2})^2 \, +\, \frac{(\widehat{\Gamma}^{(0)}_{11}+\widehat{\Gamma}^{(0)}_{22})^2 \, \Im[(\widehat{h}^\dag \widehat{h})_{12}]^2}{4 \, (\widehat{h}^\dag \widehat{h})_{11} (\widehat{h}^\dag \widehat{h})_{22}}}\;,
\end{align}
where $\mat{\Gamma}^{(0)}$ is the thermal width matrix, appearing in~\eqref{5.2}, but with tree-level Yukawa couplings in $\mat{\gamma}^N_{L\Phi}$.  
Taking into account the expansion of the Universe and neglecting $2\leftrightarrow 2$ scatterings in the washout term, the rate equation for the lepton asymmetry \eqref{eq:evol_dL} can be written as~\cite{Dev:2014laa}
\begin{align}
  &\frac{\D{}{[\delta \eta^L}]}{\D{}{z}} \ = \ z^3 K_1(z) \, 
  \bigg[ - \frac{1}{3} \, {\rm K} \, \delta \eta^L \:\nonumber\\&\qquad\qquad\qquad
    + \; \frac{\;\pi^2  z}{M_N^3 K_1(z) 2 \zeta(3) H_N}\bigg( \mathrm{Im}[\widehat{\upeta}^N]_{12} \, \mathrm{Im}[\widehat{\gamma}^N_{L\Phi}]_{12} \; +\; [\widehat{\upeta}^N]_{\alpha \beta} \,[\delta \widehat{\gamma}^N_{L \Phi}]_{\beta \alpha} \bigg) \bigg] \;,
\label{5.7}
\end{align}
where $\mathrm{K} = \Tr \mathbf{K}$ is an effective washout parameter. The attractor solution is found by setting the RHS of~\eqref{5.7} to zero. We also neglect the $O(h^6)$ off-diagonal entries in the last term, finally obtaining
\begin{equation}
\delta \eta^L \ \simeq \ \delta \eta^L_{\rm mix} \; +\; \delta \eta^L_{\rm osc} \; ,
\label{5.8}
\end{equation}
where the neglected terms in~\eqref{5.8} are formally at $O(h^6)$ and higher. 
Here, the mixing and oscillation contributions are given explicitly by 
\begin{align}
\label{mix}
\delta \eta^L_{\rm mix} \ &= \ \frac{g_N}{2} \frac{3}{2\mathrm{K} z} \; \sum_{\alpha \neq \beta}\,\frac{\Im \big(\widehat{h}^\dag \widehat{h})^2_{\alpha \beta}}
  {(\widehat{h}^\dag \widehat{h})_{11}(\widehat{h}^\dag \widehat{h})_{22}}\; \frac{\big(M^2_{N,\,\alpha} - M^2_{N,\,\beta}\big)
   M_N \widehat{\Gamma}_{\beta \beta}^{(0)}}
  {\big(M^2_{N,\,\alpha} - M^2_{N,\,\beta}\big)^2
    +\big( M_N \widehat{\Gamma}_{\beta \beta}^{(0)} \big)^2} \;, \\[0.5em]
\label{osc}
\delta \eta^L_{\rm osc} \ &= \ \frac{g_N}{2} \frac{3}{2\mathrm{K} z} \; \frac{\Im \big(\widehat{h}^\dag \widehat{h})^2_{12}}
  {(\widehat{h}^\dag \widehat{h})_{11}(\widehat{h}^\dag \widehat{h})_{22}}\; \frac{\big(M^2_{N,\,1} - M^2_{N,\,2}\big)
   M_N \big(\widehat{\Gamma}_{11}^{(0)} + \widehat{\Gamma}_{22}^{(0)}\big)}
  {\big(M^2_{N,\,1} - M^2_{N,\,2}\big)^2
    \,+\, M_N^2(\widehat{\Gamma}^{(0)}_{11}+\widehat{\Gamma}^{(0)}_{22})^2 \, \frac{\Im[(\widehat{h}^\dag \widehat{h})_{12}]^2}{ (\widehat{h}^\dag \widehat{h})_{11} (\widehat{h}^\dag \widehat{h})_{22}}} \;.
\end{align}
These results, valid in the weakly-resonant strong-washout regime, exactly reproduce the ones obtained in the semi-classical Boltzmann approach of~\cite{Dev:2014laa} for the single lepton flavour case studied here. At leading order, the contribution of mixing is governed by the diagonal entries of the $\CP$-``even'' number density $\mat{\underline{\widehat{n}}}^N$, whereas that of oscillations is triggered by the presence of off-diagonal $\CP$-``odd'' $\mat{\delta \widehat{n}}^N$. A detailed discussion of both flavour mixing and oscillations, in relation to the $\CP$-violation properties of the Lagrangian of the system, is given in~\cite{Dev:2014laa}.  

We note here that the oscillation term in \eqref{osc} by itself agrees with the form for the {\em total} asymmetry given in the quantum Boltzmann approach of~\cite{Garbrecht:2014aga} and with earlier results of~\cite{Garbrecht:2011aw,Garny:2011hg,Iso:2013lba, Iso:2014afa} in their validity  limit ${\rm Re}[(\widehat{h}^\dag \widehat{h})_{12}^2]\ll (\widehat{h}^\dag \widehat{h})_{\alpha \alpha}$.\footnote{The limit ${\rm Re}[(\widehat{h}^\dag \widehat{h})_{12}^2]\ll (\widehat{h}^\dag \widehat{h})_{\alpha \alpha}$ implies that $\Im[(\widehat{h}^\dag \widehat{h})_{12}]^2\simeq (\widehat{h}^\dag \widehat{h})_{11} (\widehat{h}^\dag \widehat{h})_{22}$ in the two-flavour case.}
Moreover, this oscillation phenomenon does \emph{not} involve any off-shell effects, since \eqref{osc} can be obtained from an on-shell analysis with only tree-level Yukawa couplings (see~\cite{Dev:2014laa}). 
However, unlike previous treatments, we emphasize that the KB approach detailed in this paper captures the distinct phenomena of flavour mixing \cite{Pilaftsis:1997dr, Pilaftsis:1997jf, Pilaftsis:2003gt, Flanz:1994yx, Covi:1996wh, Buchmuller:1997yu} in addition to oscillation phenomena. As shown numerically in~\cite{Dev:2014laa}, the contributions of these two distinct flavour effects, \eqref{mix} and \eqref{osc}, are comparable in the weakly-resonant regime. Hence, the total lepton asymmetry in \eqref{5.8} can be enhanced by a factor of order two, compared to either \eqref{mix} or \eqref{osc} alone. 

\begin{figure}[t]
  \hspace{1cm}
 \includegraphics[width=13cm]{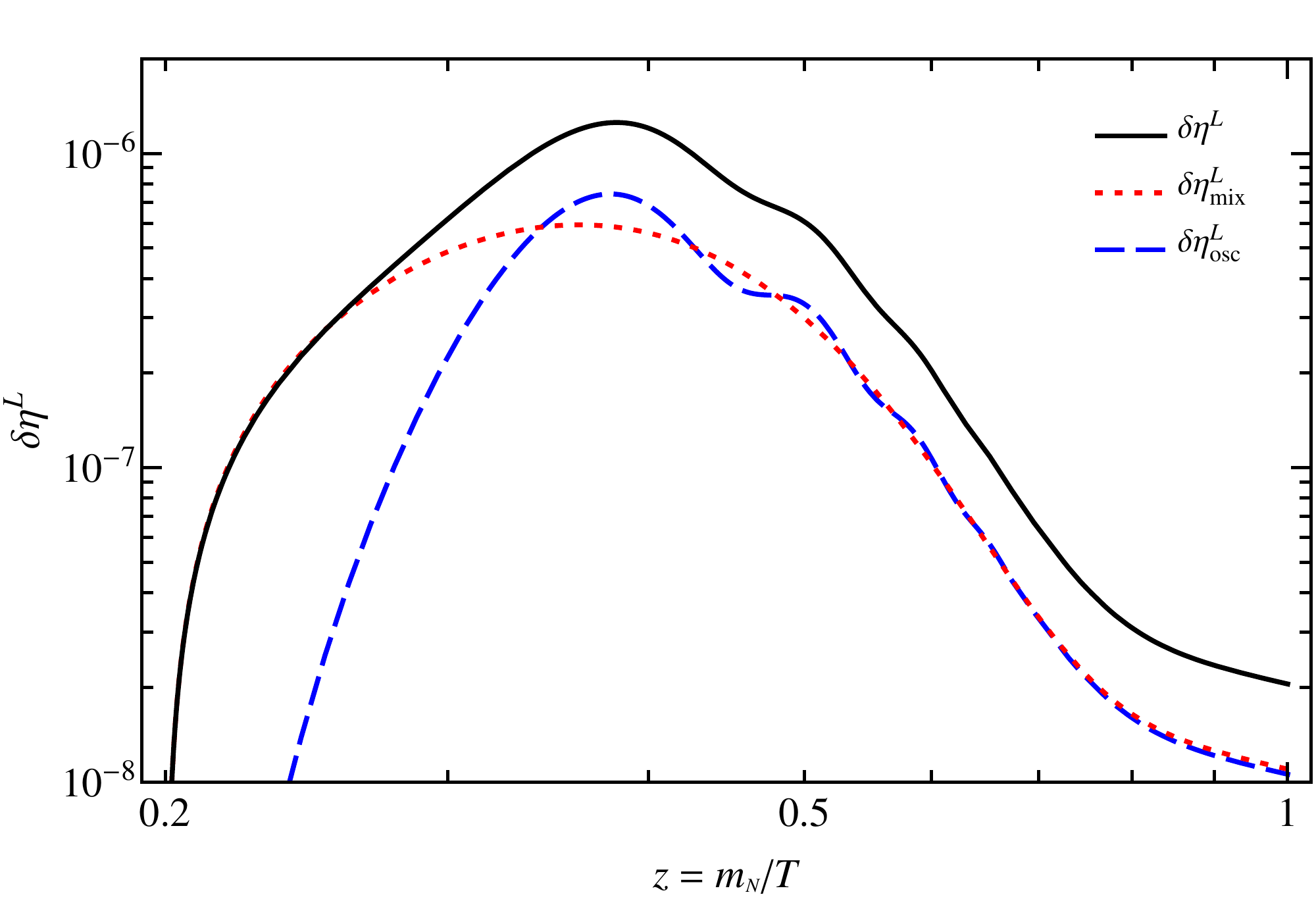}
  \caption{The evolution of the total asymmetry (black continuous line), starting from the initial conditions $\mat{\eta}^N = 2 \,\eta^N_{\rm eq} \mat{1}_2$ and $\delta \eta^L = 0$. The red dotted line is the contribution of flavour mixing and the blue dashed line is that of oscillations. For illustrative purposes, the parameters are chosen as follows: $m_N = 1 \, \mathrm{TeV}$, $(\widehat{m}_{N,\,2}-\widehat{m}_{N,\,1})/m_N = 10^{-12}$, $\widehat{h}^{1} = 0.5 \times 10^{-6} \, (1 + 5 \, e^{i \delta})\, m_N$ and $\widehat{h}^{2} = 0.5\,i \times 10^{-6} \, (1 - 5 \, e^{i \delta})\, m_N$, with $\delta = 2 \times 10^{-5}$. For simplicity, the effect of thermal masses and widths is neglected. 
\label{fig:mix_osc}}
\end{figure}

In order to illustrate the \emph{distinction} between the two physical phenomena contributing to the generation of the asymmetry, we plot in Figure~\ref{fig:mix_osc} the numerical solution of the rate equations \eqref{eq:upeta_N} and \eqref{5.7}, starting from the initial conditions $\mat{\eta}^N = 2 \, \eta^N_{\rm eq} \mat{1}_2$ and $\delta \eta^L = 0$. The black continuous line denotes the solution of the full rate equations, the red dotted line gives the contribution of mixing (obtained neglecting off-diagonal number densities) and the blue dashed line shows the contribution of oscillations (obtained replacing $\hr{}{\alpha} \to \h{}{\alpha}$).
With this choice of initial conditions, coherences between the two heavy-neutrino flavours are initially absent. Thus, at early times, only flavour mixing contributes to the asymmetry. On the other hand, as discussed in detail in \cite{Dev:2014laa}, in order to have a significant contribution from oscillations, the off-diagonal entry $n^N_{12}$ needs first to be created by coherent decays and inverse decays. Thus, as shown in Figure~\ref{fig:mix_osc}, this phenomenon becomes effective later than flavour mixing. At late times, both phenomena are present and give a similar contribution in the weakly-resonant strong-washout regime, providing an enhancement by a factor of order two with respect to mixing or oscillations alone. The different time behaviour outlined above, in addition to their differing physical origins, confirms that mixing and oscillations are two distinct physical phenomena and that both their contributions to the asymmetry need to be taken into account. Finally, we point out that the oscillatory behaviour in Figure~\ref{fig:mix_osc} does not result from non-Markovian memory effects as studied, for example, in \cite{De Simone:2007rw,De Simone:2007pa}. Instead, it is due to the oscillation of the heavy-neutrino coherences.
The non-Markovian finite-time effects are safely neglected in the strong-washout regime of interest here.

Before concluding this section, we would like to stress here that the phenomenon of coherent heavy-neutrino oscillations, discussed above, is an $O(h^4)$ effect on the {\it total} lepton asymmetry~\cite{Dev:2014laa} and so  differs from the $O(h^6)$ mechanism proposed in~\cite{Akhmedov:1998qx}. The latter effect is relevant only at temperatures much higher than the sterile neutrino masses, such as in the models studied in~\cite{Akhmedov:1998qx, Asaka:2005, Shaposhnikov:2008pf, Canetti:2012kh, Drewes:2012ma, Gagnon:2010kt, Asaka:2011wq, Shuve:2014zua, Canetti:2014dka}, where the total lepton number is not violated at leading order. On the other hand, the $O(h^4)$ effect identified here (and earlier in~\cite{Dev:2014laa}) is enhanced in the same regime as the resonant $\varepsilon$-type $ \CP$ violation effects, namely for
$z       \approx       1$       and       $\Delta       M_N       \sim
\Gamma_{N_\alpha}$, and its contribution to the final lepton asymmetry  depends crucially on the flavour coherences in the heavy-neutrino sector (cf.~\eqref{eq:evol_dL}; see~\cite{Dev:2014laa} for a detailed discussion). In the current work, we have assumed that the momentum distribution in kinetic equilibrium is a flavour singlet. As discussed in \cite{Dev:2014laa}, this approximation is valid in the resonant regime, but not in the hierarchical one. A detailed study of this phenomenon in the hierarchical regime goes beyond the scope of this paper and will be given elsewhere.
 
\section{Conclusions}
\label{sec:conc}

We have presented a novel approach to the study of flavour effects in Resonant Leptogenesis by embedding the fully flavour-covariant formalism developed in~\cite{Dev:2014laa} into the perturbative  non-equilibrium thermal field theory formulated in~\cite{Millington:2012pf}. In this formulation, one may expand the Schwinger-Dyson series diagrammatically in a perturbative loopwise sense, without encountering pinch singularities. Moreover, one may define physically-meaningful number densities at any order in perturbation theory, without necessitating the use of any quasi-particle ansatz. The truncation of the resulting transport equations proceeds in a two-fold manner: (i) spectrally, corresponding to the choice of observables being counted in the quantum transport equation and (ii) statistically, by which the set of processes causing the non-equilibrium evolution of the system are fixed.

Within this perturbative non-equilibrium field-theoretical framework, we have confirmed the results previously obtained in \cite{Dev:2014laa} via a semi-classical formalism, reproducing them quantitatively at $O(h^4)$ in the weakly-resonant regime. The main physical result is that the mixing of different heavy-neutrino flavours and the oscillations between them are two \emph{distinct} physical phenomena. The first is driven by the $\CP$-``even'' number density $\mat{\underline{n}}^N$ and the $\CP$-``odd'' rate $[\delta \gamma^N_{L\Phi}]_\alpha^{\phantom \alpha \beta}$, whereas the second is mediated by the $\CP$-``odd'' off-diagonal coherences $[\delta \widehat{n}^N]_{12}$. This is akin to the mixing and oscillation phenomena observed experimentally in $K$, $D$ and $B$-meson systems.  As identified in \cite{Dev:2014laa}, both the phenomena  contribute at $O(h^4)$ with comparable magnitude in the weakly-resonant regime. The strong-washout form of the asymmetry due to oscillations \eqref{osc} is in agreement with the results obtained in other KB studies \cite{Garbrecht:2011aw,Garny:2011hg,Iso:2013lba,Garbrecht:2014aga} and in the flavour-covariant semi-classical approach in \cite{Dev:2014laa}. 

However, as emphasized throughout this article, the KB approach presented here includes \emph{also} the effect of mixing, as given by \eqref{mix}. This contribution agrees with the one identified in \cite{Pilaftsis:1997jf}, and re-obtained in \cite{Pilaftsis:2003gt, Dev:2014laa}, once the thermal masses and widths are used in the formulae given there. The appearance of this additional $O(h^4)$ contribution, not present in previous KB studies, is due to the fact that we do not, as is typically the case, use a KB ansatz, or other equivalent approximations, for the dressed heavy-neutrino propagators. We have shown that these approximations implicitly discard mixing effects. In the approach detailed here, such approximations are not required, since we are able to express the source term for the asymmetry in terms of the spectrally-free heavy-neutrino propagators, with the effect of mixing being captured by the effective resummed Yukawa couplings [cf.~\eqref{eq:source_res_Yuk}].
In Figure~\ref{fig:mix_osc}, we have shown explicitly that mixing and oscillations are two distinct physical phenomena that contribute separately to the asymmetry, since their time behaviour, in addition to their physical origin, is different. With this approach to solving the quantum transport equations, we have justified, at leading order in the weakly-resonant regime, the semi-classical approach adopted in~\cite{Dev:2014laa} of describing the effect of mixing by means of effective $\CP$-violating Yukawa couplings \cite{Pilaftsis:2003gt}. Finally, we emphasise that mixing and oscillation contributions to the BAU are not exclusive to leptogenesis but generic phenomena applicable to baryogenesis models involving mixing of states. Therefore, both contributions should be included for precise quantitative predictions of the generated baryonic asymmetry in the Universe.

\section*{Acknowledgments}

The authors would like to thank Bj\"{o}rn Garbrecht for critical reading of the manuscript. D.T. would also like to thank Alexander Kartavtsev for discussions. The work  of P.S.B.D. and  A.P.  is supported  by the
Lancaster-Manchester-Sheffield  Consortium   for  Fundamental  Physics
under  STFC   grant  ST/J000418/1. The work of P.M. is supported by a University Foundation Fellowship (TUFF) from the Technische Universit\"{a}t M\"{u}nchen and by the DFG cluster of
excellence `Origin and Structure of the Universe'. This work was also supported in part by the IPPP under STFC grant ST/G000905/1. The work of D.T. has been supported by a fellowship of the EPS  Faculty of  the  University  of Manchester.  


\appendix

\section{Resummed Thermal Propagators and Yukawa Couplings}
\label{app:resprop}

In \cite{Dev:2014laa}, it was shown that the time-translational invariance of flavour-covariant CTP propagators is necessarily broken in the absence of thermodynamic equilibrium. In this section, working within the Markovian approximation detailed in Sections \ref{sec:source} and \ref{sec:KBansatz}, we derive the momentum-space representation of the resummed CTP propagator in the mass eigenbasis. Subsequently, in \ref{app:res_Yuk}, we use the form of these resummed propagators to obtain the resummed Yukawa couplings in the presence of thermal corrections. In so doing, we generalize the approach of \cite{Pilaftsis:2003gt}. Finally, in~\ref{app:RIS}, we reproduce the thermal RIS contribution used in the semi-classical approach of \cite{Dev:2014laa}. Throughout this appendix, we suppress the superscript~$N$ on heavy-neutrino propagators and self-energies for notational convenience.

In coordinate space, the resummed heavy-neutrino CTP propagator takes the form
\begin{equation}
\label{eq:prop1}
[i\Delta^{ab}(x,y,\tilde{t}_f;\tilde{t}_i)]_{\alpha}^{\phantom{\alpha}\beta}\ \equiv\ \braket{\mathrm{T}_{\mathcal{C}}[N_{\alpha}^a(x)N^{b,\,\beta}(y)]}_t\;,
\end{equation}
where $\mathrm{T}_{\mathcal{C}}$ denotes path ordering along the CTP contour (see Figure~\ref{fig:contour}), $a,\:b\:=\:1,\:2$ are the CTP indices (see \cite{Jordan:1986ug, Calzetta:1986ey, Calzetta:1986cq}) and the heavy-neutrino field operators are understood in the Heisenberg picture. We note that \eqref{eq:prop1} is \emph{not} a picture-independent object, since it is not evaluated at equal times $x^0=y^0=\tilde{t}_f$ (see \cite{Millington:2012pf}). In what follows, we work in momentum space, omitting all arguments for conciseness.

In the thermal mass eigenbasis (see Section~\ref{sec:model}), we may write the inverse resummed CTP propagator $\widehat{\bm{\Delta}}^{-1}$ in the following block decomposition:
\begin{equation}
\label{eq:inv}
\widehat{\bm{\Delta}}^{-1}\ =\ \begin{bmatrix} \mat{D} & -\:\mat{D}_{<} \\ -\:\mat{D}_{>} & \overline{\mat{D}} \end{bmatrix}
\end{equation}
where the set of submatrices $\{\mat{D}\,\}$ have elements given by (see e.g.~\cite{Pilaftsis:2003gt}) 
\begin{subequations}
\label{Ddefs}
\begin{align}
D_{\alpha\beta}\ &=\  (p^2\:-\:\widehat{M}_{\alpha}^2)\,\delta_{\alpha\beta}\:+\:i\epsilon(\delta_{\alpha\beta}+2 n_{\alpha\beta})\:+\:[\widehat{\Pi}_{\mathrm{abs}}]_{\alpha\beta}\;,\\
D_{\gtrless,\,\alpha\beta}\ &=\ 2i\epsilon \big(\theta(\pm\,p_0)\delta_{\alpha\beta}\:+\:n_{\alpha\beta}\big)\:+\:[\widehat{\Pi}_{\gtrless}]_{\alpha\beta}\;,\\
\overline{D}_{\alpha\beta}\ &=\ -\:(p^2\:-\:\widehat{M}_{\alpha}^2)\,\delta_{\alpha\beta}\:+\:i\epsilon(\delta_{\alpha\beta}+2n_{\alpha\beta})\:-\:[\widehat{\Pi}_{\mathrm{abs}}^{*}]_{\alpha\beta}\;.
\end{align}
\end{subequations}
Here, $\widehat{M}_\alpha$ is the thermal mass, defined via \eqref{eq:thermal_mass}, and $[\widehat{\Pi}_{\mathrm{abs}}]_{\alpha\beta}$ are the elements of the absorptive part of the Feynman self-energy. We omit the caret on $\mat{D}$'s for notational convenience. The terms proportional to the prescription $\epsilon$ (cf.~\cite{Millington:2012pf}) are, as we will see, necessary to obtain the correct tree-level propagators and results consistent with the diagrammatic resummation in Section~\ref{sec:KBansatz}. We note that the matrix inversion of the inverse propagator does not yield a unique solution to the Klein-Gordon equation without correctly encoding the boundary conditions of the Cauchy problem by virtue of these prescription-dependent terms.

The inverse CTP propagator \eqref{eq:inv} transforms as a rank-2 tensor of $U(\mathcal{N})$ under an arbitrary flavour rotation $\mathbf{U}$, as follows:
\begin{equation}
	\label{block1}
	[\bm{\Delta}^{-1}]_{k}^{\phantom{k}l}\ =\ [\bm{\mathcal{U}}^{\dag}\,\widehat{\bm{\Delta}}^{-1}\,\bm{\mathcal{U}}]_{k}^{\phantom{k} l}
\end{equation}
where $\bm{\mathcal{U}} \in U(\mathcal N)$ and can be written as a Kronecker product
\begin{equation}
\bm{\mathcal{U}}\ \equiv\ \bm{1}_2\:\otimes\: \mathbf{U}\;,
\end{equation}
in which $\bm{1}_2$ is the $2\times 2$ unit matrix.
In addition, the CTP indices of $\widehat{\bm{\Delta}}^{-1}$ in \eqref{block1} are raised and lowered by means of the $SO(1,1)$ CTP `metric'
\begin{equation}
\bm{g}\ =\ g^{ab}\ \equiv\ \mathrm{diag}(1,-1)\;,
\end{equation}	
as follows:
\begin{equation}
	[\widehat{\bm{\Delta}}^{-1}]^{ab}\ =\ [\mathbf{g}\,\widehat{\bm{\Delta}}^{-1}\,\mathbf{g}]^{ab}\;,
\end{equation}
where
\begin{equation}
\mathbf{g}\ \equiv\ \bm{g}\:\otimes\:\bm{1}_2\;.
\end{equation}
Notice that the choice of block decomposition is not unique. We could alternatively have chosen to represent the inverse CTP propagator in the form
\begin{equation}
\label{block2}
[\widehat{\bm{\Delta}}^{-1}]'\ \equiv\ \begin{bmatrix} \mat{D}_{11} & \mat{D}_{12} \\ \mat{D}_{21} & \mat{D}_{22} \end{bmatrix}\;,
\end{equation}
where
\begin{equation}
\mat{D}_{\alpha\beta}\ =\ \begin{bmatrix} D_{\alpha\beta} & -\:D_{<,\,\alpha\beta} \\ -\:D_{>,\,\alpha\beta} & \overline{D}_{\alpha\beta} \end{bmatrix}\;.
\end{equation}
In this case, the order of the Kronecker products in the $2\mathcal{N}\times 2\mathcal{N}$ $U(\mathcal{N})$ and $SO(1,1)$ transformation matrices would be reversed, i.e.
\begin{equation}
\bm{\mathcal{U}}\ \equiv\  \mathbf{U}\:\otimes\:\bm{1}_2\;,\qquad \mathbf{g}\ \equiv\ \bm{1}_2\:\otimes\:\bm{g}\;.
\end{equation}
Nevertheless, the two block decompositions \eqref{block1} and \eqref{block2} are related by means of a permutation transformation, i.e.~$[\bm{\Delta}^{-1}]'\:=\:\mathbf{P}\bm{\Delta}^{-1}\mathbf{P}$. For example, in the relevant case $\mathcal{N}=2$, the involutory permutation matrix $\mathbf{P}$ is given by
\begin{equation}
\mathbf{P}\ =\ \begin{bmatrix} 1 & 0 & 0 & 0\\ 0 & 0 & 1 & 0\\ 0 & 1 & 0 & 0\\ 0 & 0 & 0 & 1\end{bmatrix}\;.
\end{equation}
Therefore, both choices of block decomposition will yield equivalent results for the resummed CTP propagator, since these will be related by the same transformation,  i.e.~$\bm{\Delta}'\:=\:\mathbf{P}\bm{\Delta}\mathbf{P}$.

By virtue of the Banachiewicz inversion formula, the block decomposition of the resummed CTP propagator is
\begin{equation}
\bm{\Delta}\ \equiv\ [\bm{\Delta}]^{ab}\ =\ \begin{bmatrix} \big(\bm{\Delta}^{-1}/\overline{\mat{D}}\big)^{-1} & \big(\bm{\Delta}^{-1}/\mat{D}_{<}\big)^{-1} \\ \big(\bm{\Delta}^{-1}/\mat{D}_{>}\big)^{-1} & \big(\bm{\Delta}^{-1}/\mat{D}\big)^{-1} \end{bmatrix}\;,
\end{equation}
where $\mathbf{A}/\mathbf{B}$ denotes the Schur complement of $\mathbf{A}$ relative to $\mathbf{B}$, i.e.
\begin{subequations}
\begin{gather}
\bm{\Delta}^{-1}/\overline{\mat{D}}\ =\ \mat{D}\:-\:\mat{D}_<\overline{\mat{D}}^{-1}\mat{D}_>\;,\qquad
\bm{\Delta}^{-1}/\mat{D}\ =\ \overline{\mat{D}}\:-\:\mat{D}_>\mat{D}^{-1}\mat{D}_<\;,\\
\bm{\Delta}^{-1}/\mat{D}_>\ =\ \mat{D}_<\:-\:\mat{D}\mat{D}_{>}^{-1}\overline{\mat{D}}\;,\qquad
\bm{\Delta}^{-1}/\mat{D}_<\ =\ \mat{D}_>\:-\:\overline{\mat{D}}\mat{D}_{<}^{-1}\mat{D}\;.
\end{gather}
\end{subequations}
The resummed CTP propagator then takes the form
\begin{equation}
\bm{\Delta}\ =\ \frac{1}{ \mathrm{det}\,\bm{\Delta}^{-1}}\begin{bmatrix} \pmb{\mathscr{D}} & \pmb{\mathscr{D}}^< \\ \pmb{\mathscr{D}}^> & \overline{\pmb{\mathscr{D}}}\end{bmatrix}\;,
\end{equation}
where
\begin{gather}
\pmb{\mathscr{D}}\ \equiv\ \overline{\mathrm{D}}\, \mathrm{adj}\big(\mat{D}\:-\:\mat{D}_{<}\overline{\mat{D}}^{-1}\mat{D}_>\big)\;,\qquad \overline{\pmb{\mathscr{D}}}\ \equiv\ \mathrm{D}\, \mathrm{adj}\big(\mat{D}\:-\:\mat{D}_{>}\mat{D}^{-1}\mat{D}_<\big)\;,\\
\pmb{\mathscr{D}}_>\ \equiv\ \mathrm{D}_>\, \mathrm{adj}\big(\mat{D}_<\:-\:\mat{D}\mat{D}_{>}^{-1}\overline{\mat{D}}\big)\;,\qquad 
\pmb{\mathscr{D}}_<\ \equiv\ \mathrm{D}_<\, \mathrm{adj}\big(\mat{D}_>\:-\:\overline{\mat{D}}\mat{D}_{<}^{-1}\mat{D}\big)\;.
\end{gather}
Here, $\mathrm{adj}$ indicates the adjugate matrix and Roman (non-italicized) $\mathrm{D}$'s denote the determinant of the corresponding matrix, e.g.~for $\mathcal{N}=2$
\begin{equation}
\mathrm{D}\ =\ \mathrm{det}\,\mat{D}\ =\ D_{11}D_{22}\:-\:D_{12}D_{21}\;.
\end{equation}
Using the relations for the retarded and advanced functions,
\begin{subequations}
\begin{gather}
\mat{D}_{\mathrm{R}}\ =\ \mat{D}\:-\:\mat{D}_<\ =\ \mat{D}_>\:-\:\overline{\mat{D}}\;,\\
\mat{D}_{\mathrm{A}}\ =\ \mat{D}_{\mathrm{R}}^{\dag}\ =\ \mat{D}\:-\:\mat{D}_>\ =\ \mat{D}_<\:-\:\overline{\mat{D}}\;,
\end{gather}
\end{subequations}
we may show that
\begin{subequations}
\begin{align}
\pmb{\mathscr{D}}\  &=\ -\:\mathrm{adj}\,(\mat{D}_{\mathrm{R}})\,\overline{\mat{D}}\,\mathrm{adj}\,(\mat{D}_{\mathrm{A}})\;,\\
\overline{\pmb{\mathscr{D}}}\  &=\ -\:\mathrm{adj}\,(\mat{D}_{\mathrm{R}})\,\mat{D}\,\mathrm{adj}\,(\mat{D}_{\mathrm{A}})\;,\\
\pmb{\mathscr{D}}_{\gtrless}\  &=\ -\:\mathrm{adj}\,(\mat{D}_{\mathrm{R}})\,\mat{D}_{\gtrless}\,\mathrm{adj}\,(\mat{D}_{\mathrm{A}})\;.
\end{align}
\end{subequations}

The determinant of the inverse CTP propagator may be calculated using elementary row transformations and is given by
\begin{equation}
\mathrm{det}\,\bm{\Delta}^{-1} =\ (-1)^{\mathcal{N}}\:\mathrm{D}_{\mathrm{R}}\mathrm{D}_{\mathrm{A}}\ =\ (-1)^{\mathcal{N}}\mathrm{D}_{\mathrm{R}}\mathrm{D}_{\mathrm{R}}^{*}\ =\ (-1)^{\mathcal{N}}|\mathrm{D}_{\mathrm{R}}|^2.
\end{equation}

Finally, putting everything back together, we find the following form for the resummed CTP propagators in the case of $\mathcal{N}$ flavours:
\begin{subequations}
\label{resprops}
\begin{align}
\bm{\Delta}_{\mathrm{F}}\ &=\ (-1)^{\mathcal{N}-1}\bm{\Delta}_{\mathrm{R}}\,\overline{\mat{D}}\,\bm{\Delta}_{\mathrm{A}}\;,\\
\bm{\Delta}_{\mathrm{D}}\ &=\ (-1)^{\mathcal{N}-1}\bm{\Delta}_{\mathrm{R}}\,\mat{D}\,\bm{\Delta}_{\mathrm{A}}\;,\\
\bm{\Delta}_{\gtrless}\ &=\ (-1)^{\mathcal{N}-1}\bm{\Delta}_{\mathrm{R}}\,\mat{D}_{\gtrless}\,\bm{\Delta}_{\mathrm{A}}\;,
\end{align}
\end{subequations}
where
\begin{equation}
\bm{\Delta}_{\mathrm{R}}\ =\ \mat{D}_{\mathrm{R}}^{-1}\ =\ \frac{\mathrm{adj}\,\mat{D}_{\mathrm{R}}}{\mathrm{D}_{\mathrm{R}}}\;,\qquad \bm{\Delta}_{\mathrm{A}}\ =\ \mat{D}_{\mathrm{A}}^{-1}\ =\ \frac{\mathrm{adj}\,\mat{D}_{\mathrm{A}}}{\mathrm{D}_{\mathrm{A}}}
\end{equation}
are the retarded and advanced propagators, respectively. We note that the expressions \eqref{resprops} are \emph{fully} flavour-covariant and can be rotated to \emph{any} basis. In addition, one may verify that
\begin{equation}
\bm{\Delta}_{\mathrm{R}(\mathrm{A})}\ =\ \bm{\Delta}_{\mathrm{F}}\:-\:\bm{\Delta}_{<(>)}\ =\ \bm{\Delta}_{>(<)}\:-\:\bm{\Delta}_{\mathrm{D}}\;,
\end{equation}
consistent with a single-flavour scenario.

In addition, we note that by virtue of the contributions from the $\epsilon$-dependent terms in~\eqref{Ddefs}, the resummed propagators in \eqref{resprops} are consistent with those obtained by the iterative diagrammatic resummation in \eqref{fullWight}. For instance, consider the $\epsilon$-dependent contribution to the Wightman propagators
\begin{equation}
i[\widehat{\Delta}_{\gtrless}]_{\alpha\beta}\ \supset\ [\widehat{\Delta}_{\mathrm{R}}]_{\alpha\gamma} 2\epsilon \big(\theta(\pm p_0)\delta_{\gamma\delta}\:+\:\widehat{n}_{\gamma\delta}\big)[\widehat{\Delta}_{\mathrm{A}}]_{\delta \beta}\;.
\end{equation}
This may be written as
\begin{equation}
\label{gtrlessmid}
i[\widehat{\Delta}_{\gtrless}]_{\alpha\beta}\ \supset\ [\widehat{\Delta}_{\mathrm{R}}]_{\alpha\gamma}[\widehat{\Delta}_{\mathrm{R}}^{0,\,-1}]_{\gamma\sigma}[\widehat{\Delta}^0_{\mathrm{R}}]_{\sigma\sigma} 2\epsilon [\theta(\pm p_0)\delta_{\sigma\rho}\:+\:\widehat{n}_{\sigma\rho}][\widehat{\Delta}_{\mathrm{A}}^0]_{\rho\rho}[\widehat{\Delta}_{\mathrm{A}}^{0,\,-1}]_{\rho\delta}[\widehat{\Delta}_{\mathrm{A}}]_{\delta\beta}\;,
\end{equation}
where the central terms are given by 
\begin{equation}
[\widehat{\Delta}^0_{\mathrm{R}}]_{\sigma\sigma} 2\epsilon [\theta(\pm p_0)\delta_{\sigma\rho}\:+\:n_{\sigma\rho}][\widehat{\Delta}_{\mathrm{A}}^0]_{\rho\rho}\ =\ \frac{1}{p^2-\widehat{M}_{\sigma}+i\epsilon p_0}2\epsilon [\theta(\pm p_0)\delta_{\sigma\rho}\:+\:\widehat{n}_{\sigma\rho}]\frac{1}{p^2-\widehat{M}_{\rho}-i\epsilon p_0}\;.
\end{equation}
In the homogeneous Markovian approximation, we replace $\widehat{M}_{\sigma}\sim \widehat{M}_{\rho} \approx \widehat{M}$. Thus, using the limit representation of the Dirac delta function
\begin{equation}
\delta(x)\ =\ \lim_{\epsilon\to 0^+}\frac{1}{\pi}\frac{\epsilon}{x^2+\epsilon^2}\;,
\end{equation}
we find
\begin{equation}\label{eq:prop_epsilon}
i[\widehat{\Delta}^0_{\gtrless}]_{\sigma\rho}\ =\ [\widehat{\Delta}^0_{\mathrm{R}}]_{\sigma\sigma} 2\epsilon [\theta(\pm p_0)\delta_{\sigma\rho}\:+\:\widehat{n}_{\sigma\rho}][\widehat{\Delta}_{\mathrm{A}}^0]_{\rho\rho}\ =\ 2\pi\delta(p^2-\widehat{M}^2)[\theta(\pm p_0)\delta_{\sigma\rho}\:+\:\widehat{n}_{\sigma\rho}]\;,
\end{equation}
which is precisely the propagator in \eqref{homogenN}. Hence, \eqref{gtrlessmid} yields the second line of \eqref{fullWight}. 

Having observed that it is not appropriate to neglect the $\epsilon$-dependent terms next to the self-energies in \eqref{Ddefs}, it is pertinent to comment on the NWA of the resummed propagators. At first sight, it would appear that both lines of \eqref{fullWight} give two identical contributions in the NWA. However, we point out that $\epsilon$ and $\eta \equiv \Im \Pi_R \to 0$ should be treated as two independent infinitesimals, since the latter is, strictly speaking, small but finite in the NWA. Thus, in this approximation, the first line of \eqref{fullWight} is proportional to $\eta$, whereas the second line to $\epsilon$ [see \eqref{eq:prop_epsilon}]. Combining them, we obtain a term of the form
\begin{equation}
\lim_{\epsilon, \eta\to 0^+} \frac{1}{\pi} \, \frac{\epsilon + \eta}{x^2+(\epsilon + \eta)^2}\ =\ \delta(x)\;,
\end{equation}
which shows that we recover the expected result in the NWA.

In addition, it is illustrative to check that we recover the correct zero-temperature and single-flavour CTP limits. As an example, we consider the flavour-$11$ component of the resummed Feynman propagator. In the zero-temperature limit, we may restrict to positive frequencies, setting $\Delta_<=0$, such that $\Delta_{\mathrm{R}} \to \Delta_{\mathrm{F}}$ and $\Delta_{\mathrm{A}} \to -\:\Delta_{\mathrm{D}}$. For $\mathcal{N}=2$, we then find
\begin{equation}
\label{eq:zerotemp}
[\Delta_{\mathrm{F}}]_{11}\ =\ \frac{D_{22}}{\mathrm{D}}\ =\ \Big[D_{11}\:-\:D_{12}D_{22}^{-1}D_{21}\Big]^{-1}\,
\end{equation}
which, in the mass eigenbasis, gives
\begin{equation}
[\widehat{\Delta}_{\mathrm{F}}]_{11}\ =\ \bigg[p^2\:-\:\widehat{M}_1^2\:+\:i\epsilon\:+\:\widehat{\Pi}_{11}\:-\:\frac{\widehat{\Pi}_{12}\widehat{\Pi}_{21}}{p^2\:-\:\widehat{M}_2^2\:+\:i\epsilon\:+\:\widehat{\Pi}_{22}}\bigg]^{-1}
\end{equation}
in agreement with well-known results (see e.g.~\cite{Pilaftsis:1997dr}).

On the other hand, we may obtain the single flavour limit by setting the off-diagonal components ($D_{12}$, $D_{21}$, $\Delta_{12}$, $\Delta_{21}$, etc.) to zero. In this case, we obtain the usual CTP resummed propagator
\begin{equation}
\Delta_{\mathrm{F}}\ =\ -\:\frac{\overline{D}_{11}}{|D_{\mathrm{R},\,11}|^2}\;,
\end{equation}
which, after dropping the redundant flavour indices, takes the form
\begin{equation}
\Delta_{\mathrm{F}}\ =\ \frac{p^2\:-\:M^2\:-\:i\mathrm{Im}\,\Pi(p)}{(p^2\:-\:M^2)^2\:+\:(\mathrm{Im}\,\Pi(p))^2}\;,
\end{equation}
with $M^2= m^2-\mathrm{Re}\,\Pi(p)$. Notice that we have safely dropped the $\epsilon$-dependent terms, again in agreement with known results (see e.g.~\cite{Millington:2012pf}).

\subsection{Resummed Yukawa Couplings in Charged-Lepton Self-Energies}
\label{app:res_Yuk}

In this subsection, we show explicitly that, formally at $O(h^4)$ in the asymmetry, the contribution of the charged-lepton self-energy to the source term can be written in terms of the resummed Yukawa couplings, as in \eqref{eq:source_res_Yuk} and illustrated in Figure~\ref{fig:self}. 

From \eqref{eq:evol_dL}, we see that the quantity of interest is
\begin{equation}\label{eq:amplit}
\mathcal{T} \ \equiv \ \h{}{\alpha} \, \del{<}{}{\alpha}{\beta} \, \hs{}{\beta} \;.
\end{equation}
The contribution to $\mathcal{T}$ of the second line of \eqref{fullWight}, which appears in the source term \eqref{eq:evol_dL}, will be denoted by $\mathcal{T}_{\mathrm{src}}$. Using \eqref{fullWight} and noting that the summations there are equal to $\mat{\Delta}_{\mathrm{R}(\mathrm{A})} \cdot \mat{\Delta}_{\mathrm{R}(\mathrm{A})}^{0,\,-1}$, $\mathcal{T}_{\mathrm{src}}$ can be written as
\begin{align}
\mathcal{T}_{\mathrm{src}} \ &= \ \h{}{\alpha} \, \del{\mathrm{R}}{}{\alpha}{\lambda} \, \del{\mathrm{R}}{0,\,-1}{\lambda}{\gamma} \, \del{<}{0}{\gamma}{\delta} \, \del{\mathrm{A}}{0,\,-1}{\delta}{\mu} \, \del{\mathrm{A}}{}{\mu}{\beta} \, \hs{}{\beta} \notag \\
& = \ \sum_{\gamma, \delta} \widehat{h}^{\alpha} \, \delh{\mathrm{R}}{}{\alpha}{\gamma} \, \Big(\delh{\mathrm{R}}{0,\,-1}{\gamma}{\gamma} \, \delh{<}{0}{\gamma}{\delta} \, \delh{\mathrm{A}}{0,\,-1}{\delta}{\delta} \Big) \, \delh{\mathrm{A}}{}{\delta}{\beta} \, \widehat{h}_\beta \notag \\
& \equiv \ \widehat{h}^{\alpha} \, \delh{\mathrm{R}}{}{\alpha}{\gamma} \, \widehat{\mathcal{N}}_{\gamma \delta} \, \delh{\mathrm{A}}{}{\delta}{\beta} \, \widehat{h}_\beta \;.
\end{align}
Let us introduce the notation
\begin{equation}
\slashed{\alpha} \ \equiv \ \left\{ \begin{aligned}2 \quad \text{if } \alpha = 1\\ 1 \quad \text{if } \alpha = 2
 \end{aligned}\right. \;.
\end{equation}
We treat the off-diagonal number densities $[\widehat{n}^N]_{\alpha \slashed{\alpha}}$ as formally at $O(h^2)$, assuming that they are generated dynamically from an incoherent initial condition (see \cite{Dev:2014laa}). Therefore, we have
\begin{equation}\label{eq:curly_N_expansion}
\widehat{\mathcal{N}}_{\alpha \beta} \ = \ (s - M^2_{N,\,\alpha}) \, \delh{<}{0}{\alpha}{\beta} \, (s - M^2_{N,\,\beta}) \ = \ (s - s_\alpha) \, \delh{<}{0}{\alpha}{\beta} \, (s-s_{\beta}^*) \; + \; O(h^4) \;,
\end{equation}
where we have used $\Gamma_{\alpha,\slashed{\alpha}} \delh{<}{0}{\alpha}{\slashed {\alpha}} = O(h^4)$, with $\Gamma_\alpha$ being the width of the heavy neutrinos. Here, $s_\alpha$ denotes the location of the two complex poles of the retarded propagator
\begin{equation}\label{eq:poles}
s_\alpha \ = \ M_{N,\,\alpha}^2 \, - \, i M_N \Gamma_\alpha \;.
\end{equation}
Proceeding as in \cite{Pilaftsis:2003gt}, the resonant terms in $\mathcal{T}_{\mathrm{src}}$ can be expanded as 
\begin{equation}\label{eq:fact_1}
\mathcal{T}_{\mathrm{src}} \ \simeq \ \sum_\alpha \, \widehat{\mathbf{h}}^\alpha \, \frac{Z_{\alpha}}{s - s_\alpha} \, \bigg(G_\alpha \, - \, \frac{D^{\mathrm{R}}_{12}}{D^{\mathrm{R}}_{\slashed{\alpha} \slashed{\alpha}}} \, G_{\slashed{\alpha}} \bigg) \;,
\end{equation}
with
\begin{equation}\label{eq:expans}
G_\alpha \ \equiv \ \widehat{\mathcal{N}}_{\alpha \delta} \, \delh{\mathrm{A}}{}{\delta}{\beta} \, \widehat{h}_\beta \;.
\end{equation}
In \eqref{eq:fact_1}, we have included the wavefunction renormalization $Z_{\alpha} \equiv \big(\frac{\D{}{}}{\D{}{s}} [\Delta_R(s)]_{\alpha \alpha}^{-1}\big)^{-1}$, even though this will be a higher-order effect in the analysis below. 

For the case of two heavy neutrinos studied here, the resummed Yukawa couplings are given, in the mass eigenbasis, by
\begin{equation}
\label{resum}
\widehat{\mathbf{h}}^\alpha \ = \ \widehat{h}^\alpha \; - \; \frac{\widehat{h}^{\slashed{\alpha}} \, i \, \Im [\widehat{\Pi}_{\mathrm{R}}]_{\alpha \slashed{\alpha}}}{M^2_{N,\,\alpha} - M^2_{N,\,\slashed{\alpha}} \,+\, i \, \Im [\widehat{\Pi}_{\mathrm{R}}]_{\slashed{\alpha} \slashed{\alpha}} } \;,
\end{equation}
where the indices are {\em not} summed over. The $\gCP$ conjugate couplings $\hrc{}{\alpha}$ are obtained by using the complex-conjugate tree-level couplings in the RHS of~\eqref{resum}. Equation~\eqref{eq:expans} can, in turn, be expanded as
\begin{equation}\label{eq:fact_2}
G_\alpha \ \simeq \ \sum_\beta \, \bigg( \widehat{\mathcal{N}}_{\alpha \beta} \, + \, \frac{D^{\mathrm{R}}_{12}}{D^{\mathrm{R},\,*}_{\slashed{\beta}\slashed{\beta}}} \, \widehat{\mathcal{N}}_{\alpha \slashed\beta}\bigg) \, \frac{Z_\beta^*}{s - s_\beta^*} \, \widehat{\mathbf{h}}_\beta \;.
\end{equation}
Using \eqref{eq:fact_2} in \eqref{eq:fact_1}, we find
\begin{align}
\label{A40}
\mathcal{T}_{\mathrm{src}} \ &= \ \sum_{\alpha, \beta} \, \widehat{\mathbf{h}}^\alpha \, \frac{Z_\alpha}{s - s_\alpha} \, \widehat{\mathcal{N}}_{\alpha \beta}\, \frac{Z_\beta^*}{s - s_\beta^*} \, \widehat{\mathbf{h}}_\beta  \\
&+ \ \sum_{\alpha, \beta} \,\frac{D^{\mathrm{R}}_{12}}{D^{\mathrm{R},\,*}_{\slashed{\beta}\slashed{\beta}}} \, \widehat{\mathbf{h}}^\alpha \, \frac{Z_\alpha}{s - s_\alpha} \, \widehat{\mathcal{N}}_{\alpha \slashed\beta}\, \frac{Z_\beta^*}{s - s_\beta^*} \, \widehat{\mathbf{h}}_\beta \; - \; \sum_{\alpha, \beta} \,\frac{D^{\mathrm{R}}_{12}}{D^{\mathrm{R}}_{\slashed{\alpha}\slashed{\alpha}}} \, \widehat{\mathbf{h}}^\alpha \, \frac{Z_\alpha}{s - s_\alpha} \, \widehat{\mathcal{N}}_{\slashed{\alpha} \beta}\, \frac{Z_\beta^*}{s - s_\beta^*} \, \widehat{\mathbf{h}}_\beta  \; + \; O(h^6) \;.\notag
\end{align}
The contributions in the second line of \eqref{A40} can be neglected. To show this, consider for example the first summation: the only terms that can give contributions at $O(h^4)$ are the ones with $\alpha = \slashed{\beta}$. Using \eqref{eq:curly_N_expansion}, these become
\begin{equation}
\frac{D^{\mathrm{R}}_{12}}{D^{\mathrm{R},\,*}_{\slashed{\beta}\slashed{\beta}}} \, \widehat{h}^\alpha \, \delh{<}{0}{\alpha}{\alpha}\, \frac{s - s_\alpha}{s - s_\beta^*} \, \widehat{h}_\beta \, + \, O(h^6) \ = \ \frac{D^{\mathrm{R}}_{12}}{D^{\mathrm{R},\,*}_{\slashed{\beta}\slashed{\beta}}} \,  \frac{\widehat{h}^\alpha \, \delh{<}{0}{\alpha}{\alpha}\, i M_N \Gamma_\alpha\, \widehat{h}_\beta}{M^2_{N,\,\alpha} - M^2_{N,\,\beta} - i M_N \Gamma_\beta}  \, + \, O(h^6) \ = \ O(h^6) \;,
\end{equation}
having also used \eqref{homogenN} and \eqref{eq:poles}. Therefore, we obtain the final expression
\begin{equation}
\mathcal{T}_{\mathrm{src}} \ = \ \sum_{\alpha, \beta} \, \hr{}{\alpha} \, \del{<}{0}{\alpha}{\beta} \, \hrs{}{\beta} \; + \; O(h^6) \;,
\end{equation}
which is the form that we use in \eqref{eq:source_res_Yuk} and Figure~\ref{fig:self}.

\subsection{RIS in Semiclassical Boltzmann Approaches}
\label{app:RIS}
We now use the results obtained above to recover the thermal RIS contribution used in \cite{Dev:2014laa}, relevant to semi-classical approaches. Let us consider the pole expansion of the Feynman propagator, which may be written in the form
\begin{equation}
\widehat{\bm{\Delta}}_{\mathrm{F}}\ =\ \frac{\pmb{\mathscr{D}}}{|\mathrm{D}_{\mathrm{R}}|^2}\;.
\end{equation}
For the flavour-$11$ component, the pole expansion is
\begin{equation}
[\widehat{\Delta}_{\mathrm{F}}]_{11}\ =\ \frac{\mathscr{D}_{11}}{|\mathrm{D}_{\mathrm{R}}|^2}\bigg|_{s\approx s_1}\:+\:\frac{\mathscr{D}_{11}}{|\mathrm{D}_{\mathrm{R}}|^2}\bigg|_{s\approx s_1^*}\:+\:\frac{\mathscr{D}_{11}}{|\mathrm{D}_{\mathrm{R}}|^2}\bigg|_{s\approx s_2}\:+\:\frac{\mathscr{D}_{11}}{|\mathrm{D}_{\mathrm{R}}|^2}\bigg|_{s\approx s_2^*}\:+\:\cdots\;,
\end{equation}
where $s_{1,2}^{(*)}$ are the complex roots of $|\mathrm{D}_{\mathrm{R}}|^2 =0$. Noting that
\begin{equation}
\mathrm{det}\,\pmb{\mathscr{D}}\ =\ -\:\mathrm{det}\big[\mathrm{adj}(\mat{D}_{\mathrm{R}})\overline{\mathbf{D}}\,\mathrm{adj}(\mat{D}_{\mathrm{A}})\big]\ =\ -\: |\mat{D}_{\mathrm{R}}|^{2(\mathcal{N}-1)}\mathrm{det}\,\overline{\mat{D}}\;,
\end{equation} 
it follows, in the vicinity of the poles, that
\begin{equation}
\mathrm{det}\,\pmb{\mathscr{D}}\ =\ \mathscr{D}_{11}\mathscr{D}_{22}\:-\:\mathscr{D}_{12}\mathscr{D}_{21}\ \approx\ 0\;.
\end{equation}
Hence, we may write
\begin{equation}
[\widehat{\Delta}_{\mathrm{F}}]_{11}\ =\ \frac{\mathscr{D}_{11}}{|\mathrm{D}_{\mathrm{R}}|^2}\bigg|_{s\approx s_1}\:+\:\frac{\mathscr{D}_{11}}{|\mathrm{D}_{\mathrm{R}}|^2}\bigg|_{s\approx s_1^*}\:+\:\frac{\mathscr{D}_{12}}{\mathscr{D}_{22}}\frac{\mathscr{D}_{22}}{|\mathrm{D}_{\mathrm{R}}|^2}\frac{\mathscr{D}_{21}}{\mathscr{D}_{22}}\bigg|_{s\approx s_2}\:+\:\frac{\mathscr{D}_{12}}{\mathscr{D}_{22}}\frac{\mathscr{D}_{22}}{|\mathrm{D}_{\mathrm{R}}|^2}\frac{\mathscr{D}_{21}}{\mathscr{D}_{22}}\bigg|_{s\approx s_2^*}\:+\:\cdots\;,
\end{equation}
or, equivalently,
\begin{equation}
\label{eq:res}
[\widehat{\Delta}_{\mathrm{F}}]_{11}\ =\ \frac{Z_{\mathrm{R},\,1}}{s-s_1}\:+\:\frac{Z_{\mathrm{A},\,1}}{s-s_1^*}\:+\:\frac{\mathscr{D}_{12}}{\mathscr{D}_{22}}\frac{Z_{\mathrm{R},\,2}}{s-s_2}\frac{\mathscr{D}_{21}}{\mathscr{D}_{22}}\:+\:\frac{\mathscr{D}_{12}}{\mathscr{D}_{22}}\frac{Z_{\mathrm{A},\,2}}{s-s_2^*}\frac{\mathscr{D}_{21}}{\mathscr{D}_{22}}\:+\:\cdots\;,
\end{equation}
where we have introduced
\begin{equation}
Z_{\mathrm{R}(\mathrm{A}),\,\alpha}\ \equiv\ [Z_{\mathrm{R}(\mathrm{A})}(\sqrt{s})]_{\alpha}\ =\ \bigg(\frac{\mathrm{D}_{\mathrm{A}(\mathrm{R})}(\sqrt{s})}{2\sqrt{s}}\frac{\mathrm{d}}{\mathrm{d}\sqrt{s}}\frac{[\widehat{\Delta}_{\mathrm{F}}^{-1}(\sqrt{s})]_{\alpha\alpha}}{\mathrm{D}_{\mathrm{A}(\mathrm{R})}(\sqrt{s})}\bigg)^{-1}\;.
\end{equation}
The pole expansion \eqref{eq:res} differs from that in \cite{Pilaftsis:2003gt} by the presence of the complex-conjugate poles. Proceeding similarly, we find
\begin{align}
\label{eq:res2}
[\widehat{\Delta}_{\mathrm{F}}]_{22}\ &=\ \frac{Z_{\mathrm{R},\,2}}{s-s_2}\:+\:\frac{Z_{\mathrm{A},\,2}}{s-s_2^*}\:+\:\frac{\mathscr{D}_{21}}{\mathscr{D}_{11}}\frac{Z_{\mathrm{R},\,1}}{s-s_1}\frac{\mathscr{D}_{12}}{\mathscr{D}_{11}}\:+\:\frac{\mathscr{D}_{21}}{\mathscr{D}_{11}}\frac{Z_{\mathrm{A},\,1}}{s-s_1^*}\frac{\mathscr{D}_{12}}{\mathscr{D}_{11}}\:+\:\cdots\;,\\
[\widehat{\Delta}_{\mathrm{F}}]_{12}\ &=\ \frac{Z_{\mathrm{R},\,1}}{s-s_1}\frac{\mathscr{D}_{12}}{\mathscr{D}_{11}}\:+\:\frac{Z_{\mathrm{A},\,1}}{s-s_1^*}\frac{\mathscr{D}_{12}}{\mathscr{D}_{11}}\:+\:\frac{\mathscr{D}_{12}}{\mathscr{D}_{22}}\frac{Z_{\mathrm{R},\,2}}{s-s_2}\:+\:\frac{\mathscr{D}_{12}}{\mathscr{D}_{22}}\frac{Z_{\mathrm{A},\,2}}{s-s_2^*}\:+\:\cdots\;,\\
[\widehat{\Delta}_{\mathrm{F}}]_{21}\ &=\ \frac{Z_{\mathrm{R},\,2}}{s-s_2}\frac{\mathscr{D}_{21}}{\mathscr{D}_{22}}\:+\:\frac{Z_{\mathrm{A},\,2}}{s-s_2^*}\frac{\mathscr{D}_{21}}{\mathscr{D}_{22}}\:+\:\frac{\mathscr{D}_{21}}{\mathscr{D}_{11}}\frac{Z_{\mathrm{R},\,1}}{s-s_1}\:+\:\frac{\mathscr{D}_{21}}{\mathscr{D}_{11}}\frac{Z_{\mathrm{A},\,1}}{s-s_1^*}\:+\:\cdots \; .
\end{align}

Finally, the amplitude pertinent to the derivation of the resummed Yukawa couplings in semi-classical approaches is the $s$-channel exchange~\cite{Pilaftsis:2003gt}~\footnote{Here, we have used $\mathcal{A}$ and $\mathcal{B}$ to label the vertices to avoid confusion with the $\mathrm{A}$ that denotes advanced functions. }
\begin{equation}
\mathcal{T}_s\ =\ \varGamma_1^{\mathcal{A}}\,[\widehat{\Delta}_F]_{11}\,\varGamma_1^{\mathcal{B}}\:+\:\varGamma_1^{\mathcal{A}}\,[\widehat{\Delta}_F]_{12}\,\varGamma_2^{\mathcal{B}}\:+\:\varGamma_2^{\mathcal{A}}\,[\widehat{\Delta}_F]_{21}\,\varGamma_1^{\mathcal{B}}\:+\:\varGamma_2^{\mathcal{A}}\,[\widehat{\Delta}_F]_{22}\,\varGamma_2^{\mathcal{B}}\;.
\end{equation}
Using the results above, the resonant contribution takes the form
\begin{align}
\tilde{\mathcal{T}}_s\ &=\ |V^{\mathcal{A}}_1|^2\bigg[\frac{|Z_{\mathrm{R},\,1}|^2+|Z_{\mathrm{A},\,1}|^2}{|s-s_1|^2}\:+\:\frac{Z_{\mathrm{R},\,1}Z_{\mathrm{A},\,1}^*}{(s-s_1)^2}\:+\:\frac{Z_{\mathrm{R},\,1}^*Z_{\mathrm{A},\,1}}{(s-s^*_1)^2}\bigg]|V_1^{\mathcal{B}}|^2\:+\:(1\leftrightarrow 2)\;,
\end{align}
where
\begin{equation}
V_1^{\mathcal{A}(\mathcal{B})}\ =\ \varGamma_1^{\mathcal{A}(\mathcal{B})}\:+\:\frac{\mathscr{D}_{12}}{\mathscr{D}_{11}}\varGamma_2^{\mathcal{A}(\mathcal{B})}\;,\qquad V_2^{\mathcal{A}(\mathcal{B})}\ =\ \varGamma_2^{\mathcal{A}(\mathcal{B})}\:+\:\frac{\mathscr{D}_{21}}{\mathscr{D}_{22}}\varGamma_1^{\mathcal{A}(\mathcal{B})}\;.
\end{equation}

In the pole-dominance region, we find
\begin{align}
\label{poledomin}
|\tilde{\mathcal{T}}_{s,\mathrm{RIS}}|^2\ &=\ |V_1^{\mathcal{A}}|^2|V_1^{\mathcal{B}}|^2\frac{\pi}{m_1\Gamma_{\mathrm{R},\,1}}|Z_{\mathrm{R},\,1}\:-\:Z_{\mathrm{A},\,1}|^2\delta_+(s-\widehat{M}_1^2)\:+\:(1\leftrightarrow 2)\;,
\end{align}
where $M_1^2\ =\ \mathrm{Re}\,s_1\ =\ \mathrm{Re}\,\Pi_{\mathrm{R},\,11}$ and $M_1\Gamma_{\mathrm{R},\,1}\ =\ \mathrm{Im}\,s_1\ =\ \mathrm{Im}\,\Pi_{\mathrm{R},\,11}$ are the thermal masses and widths, calculated from the dispersive and absorptive parts of the retarded self-energies, respectively. In addition, $\delta_{+}(s-\widehat{M}_1^2)=\theta(\sqrt{s})\delta(s-\widehat{M}_1^2)$. Notice that \eqref{poledomin} is obtained from the results of \cite{Pilaftsis:2003gt} by replacing $|Z_{\alpha}|^2$ with $|Z_{\mathrm{R},\,\alpha}\:-\:Z_{\mathrm{A},\,\alpha}|^2$ and the masses, widths and vertices by their thermal counterparts, calculated using the retarded self-energy.

Finally, we now show that, in the equilibrium limit, we recover the thermal RIS contribution found in \cite{Dev:2014laa}. Ignoring higher order mixing terms, we have
\begin{equation}
[\widehat{\Delta}_{\mathrm{F}}]_{11}\ \approx\ \frac{s-\widehat{M}_1^2+iM\Gamma_{F,1}}{(s-s_1)(s-s_1^*)}\ =\ \frac{s-\widehat{M}_1^2+i(1+2n(\sqrt{s}))M\Gamma_{1}}{(s-s_1)(s-s_1^*)}\;,
\end{equation}
using the fluctuation-dissipation theorem to relate the time-ordered and retarded widths $\Gamma_{\mathrm{F},\,1}$ and $\Gamma_{1}$. Partial fractioning the resonant part, we obtain
\begin{equation}
[\widehat{\Delta}_{\mathrm{F}}]_{11}\ =\ \frac{iM\Gamma_{1}}{s_1-s_1^*}(1+2n(\sqrt{s}))\bigg[\frac{1}{s-s_1}\:-\:\frac{1}{s-s_1^*}\bigg]\;.
\end{equation}
Since $s_1-s_1^*=2iM\Gamma_{1}$, we see the importance of keeping track of the structure of the numerator and find
\begin{equation}
[\widehat{\Delta}_{\mathrm{F}}]_{11}\ =\ \frac{1}{2}(1+2n(\sqrt{s}))\bigg[\frac{1}{s-s_1}\:-\:\frac{1}{s-s_1^*}\bigg]\;.
\end{equation}
In this case, the residues of the poles are
\begin{equation}
Z_{\mathrm{R},\,1}\ =\ \frac{1}{2}(1+2n(\sqrt{s}))\;,\qquad Z_{\mathrm{A},\,1}\ =\ -\:\frac{1}{2}(1+2n(\sqrt{s}))\;.
\end{equation}
Hence, the RIS contribution takes the form
\begin{align}
\label{eq:thermRIS}
|\tilde{T}_{s,\mathrm{RIS}}|^2\ &=\ |V_1^{\mathcal{A}}|^2|V_1^{\mathcal{B}}|^2\frac{\pi}{M\Gamma_{1}}(1+2n(\sqrt{s}))^2\delta_+(s-\widehat{M}_1^2)\:+\:(1\leftrightarrow 2)\:+\:\cdots\;,
\end{align}
which, to leading order in the statistical factors, contains the thermal RIS contribution identified in \cite{Dev:2014laa}. As noted in Section~\ref{sec:source}, such thermal RIS contributions are not double-counted in the KB approach discussed in this article, but must be subtracted in semi-classical Boltzmann approaches, such as \cite{Dev:2014laa}. Notice finally that, when the thermal contributions are neglected, \eqref{eq:thermRIS} agrees with the results in \cite{Pilaftsis:2003gt}.


\begin{thebibliography}{999}

\bibitem{Fukugita:1986hr}
  M.~Fukugita and T.~Yanagida,
  Phys.\ Lett.\ B {\bf 174} (1986) 45.

\bibitem{seesaw1}
  P. Minkowski,
  Phys. Lett. B {\bf 67}, 421 (1977).

\bibitem{seesaw2}
  R. N. Mohapatra and G. Senjanovi\'{c}, 
  Phys. Rev. Lett. {\bf 44}, 912 (1980). 

\bibitem{seesaw4}
  M. Gell-Mann, P. Ramond and R. Slansky, 
  Conf.\ Proc.\ C {\bf 790927}, 315 (1979)
  [arXiv:1306.4669 [hep-th]]. 

\bibitem{seesaw5}
  T. Yanagida,
  in {\it Proceedings of the Workshop on Unified Theories
    and Baryon Number in the Universe},
  eds. A. Sawada and A. Sugamoto, KEK, Tsukuba (1979). 

\bibitem{seesaw6} 
  J.~Schechter and J.~W.~F.~Valle,
  Phys.\ Rev.\ D {\bf 22}, 2227 (1980).

\bibitem{pdg} 
  K.~A.~Olive {\it et al.} (Particle Data Group),
  Chin. Phys. C {\bf 38}, 090001 (2014) [{\tt http://pdg.lbl.gov/}].
  
\bibitem{Kuzmin:1985mm}
  V.~A.~Kuzmin, V.~A.~Rubakov and M.~E.~Shaposhnikov,
  Phys.\ Lett.\ B {\bf 155} (1985) 36.

\bibitem{Sakharov:1967dj}
  A.~D.~Sakharov,
  JETP Lett.\  {\bf 5} (1967) 24. 



\bibitem{Blanchet:2012bk} 
  S.~Blanchet and P.~Di Bari,
  New J.\ Phys.\  {\bf 14}, 125012 (2012)
  [arXiv:1211.0512 [hep-ph]].


\bibitem{Pilaftsis:1997dr} 
  A.~Pilaftsis,
  Nucl.\ Phys.\ B {\bf 504}, 61 (1997)
  [hep-ph/9702393].

\bibitem{Pilaftsis:1997jf} 
  A.~Pilaftsis,
  Phys.\ Rev.\ D {\bf 56}, 5431 (1997)
  [hep-ph/9707235].

\bibitem{Pilaftsis:2003gt} 
A.~Pilaftsis and T.~E.~J.~Underwood,
  Nucl.\ Phys.\ B {\bf 692}, 303 (2004)
  [hep-ph/0309342]. 

\bibitem{Flanz:1994yx} 
  M.~Flanz, E.~A.~Paschos and U.~Sarkar,
  Phys.\ Lett.\ B {\bf 345}, 248 (1995)
  [Erratum-ibid.\ B {\bf 382}, 447 (1996)] 
  [hep-ph/9411366].

\bibitem{Covi:1996wh} 
  L.~Covi, E.~Roulet and F.~Vissani,
  Phys.\ Lett.\ B {\bf 384}, 169 (1996) 
  [hep-ph/9605319].

\bibitem{Buchmuller:1997yu} 
  W.~Buchm\"uller and M.~Pl\"umacher,
  Phys.\ Lett.\ B {\bf 431}, 354 (1998)
  [hep-ph/9710460].

\bibitem{Pilaftsis:2005rv} 
  A.~Pilaftsis and T.~E.~J.~Underwood,
  Phys.\ Rev.\ D {\bf 72}, 113001 (2005)
  [hep-ph/0506107].

\bibitem{Pilaftsis:2004xx} 
  A.~Pilaftsis,
  Phys.\ Rev.\ Lett.\  {\bf 95}, 081602 (2005)
  [hep-ph/0408103].

\bibitem{Deppisch:2010fr} 
F.~F.~Deppisch and A.~Pilaftsis,
  Phys.\ Rev.\ D {\bf 83}, 076007 (2011)
  [arXiv:1012.1834 [hep-ph]].

\bibitem{Datta:1993nm} 
  A.~Datta, M.~Guchait and A.~Pilaftsis,
  Phys.\ Rev.\ D {\bf 50}, 3195 (1994)
  [hep-ph/9311257].

\bibitem{Han:2006ip} 
  T.~Han and B.~Zhang,
  Phys.\ Rev.\ Lett.\  {\bf 97}, 171804 (2006)
  [hep-ph/0604064].

\bibitem{Bray:2007ru} 
  S.~Bray, J.~S.~Lee and A.~Pilaftsis,
  Nucl.\ Phys.\ B {\bf 786}, 95 (2007)
  [hep-ph/0702294 [HEP-PH]].

\bibitem{Atre:2009rg} 
  A.~Atre, T.~Han, S.~Pascoli and B.~Zhang,
  JHEP {\bf 0905}, 030 (2009)
  [arXiv:0901.3589 [hep-ph]].

\bibitem{Dev:2013wba} 
  P.~S.~B.~Dev, A.~Pilaftsis and U.~K.~Yang,
  Phys. Rev. Lett. {\bf 112}, 081801 (2014) [arXiv:1308.2209 [hep-ph]].
  
\bibitem{Bambhaniya:2014kga} 
  G.~Bambhaniya, S.~Goswami, S.~Khan, P.~Konar and T.~Mondal,
  arXiv:1410.5687 [hep-ph].

\bibitem{Ilakovac:1994kj} 
  A.~Ilakovac and A.~Pilaftsis,
  Nucl.\ Phys.\ B {\bf 437}, 491 (1995)
  [hep-ph/9403398].

\bibitem{Alonso:2012ji} 
  R.~Alonso, M.~Dhen, M.~B.~Gavela and T.~Hambye,
  JHEP {\bf 1301}, 118 (2013)
  [arXiv:1209.2679 [hep-ph]].


\bibitem{Pilaftsis:1998pd} 
  A.~Pilaftsis,
  Int.\ J.\ Mod.\ Phys.\ A {\bf 14}, 1811 (1999)
  [hep-ph/9812256].

\bibitem{Endoh:2003mz} 
  T.~Endoh, T.~Morozumi and Z.~-h.~Xiong,
  Prog.\ Theor.\ Phys.\  {\bf 111}, 123 (2004)
  [hep-ph/0308276].

\bibitem{Vives:2005ra} 
  O.~Vives,
  Phys.\ Rev.\ D {\bf 73}, 073006 (2006)
  [hep-ph/0512160].

\bibitem{Blanchet:2011xq} 
  S.~Blanchet, P.~Di Bari, D.~A.~Jones and L.~Marzola,
  JCAP {\bf 1301}, 041 (2013)
  [arXiv:1112.4528 [hep-ph]].
  
\bibitem{Asaka:2005} 
  T.~Asaka and M.~Shaposhnikov,
  Phys.\ Lett.\ B {\bf 620}, 17 (2005)
  [hep-ph/0505013].
  
\bibitem{Ellis:2002eh}
  J.~R.~Ellis, M.~Raidal and T.~Yanagida,
  Phys.\ Lett.\ B {\bf 546} (2002) 228
  [hep-ph/0206300].

\bibitem{Barbieri:1999ma} 
  R.~Barbieri, P.~Creminelli, A.~Strumia and N.~Tetradis,
  Nucl.\ Phys.\ B {\bf 575}, 61 (2000)
  [hep-ph/9911315].

\bibitem{Abada:2006fw} 
  A.~Abada, S.~Davidson, F.~-X.~Josse-Michaux, M.~Losada and A.~Riotto,
  JCAP {\bf 0604}, 004 (2006)
  [hep-ph/0601083].

\bibitem{Nardi:2006fx} 
  E.~Nardi, Y.~Nir, E.~Roulet and J.~Racker,
  JHEP {\bf 0601}, 164 (2006)
  [hep-ph/0601084].

\bibitem{Abada:2006ea} 
  A.~Abada, S.~Davidson, A.~Ibarra, F.~-X.~Josse-Michaux,
  M.~Losada and A.~Riotto,
  JHEP {\bf 0609}, 010 (2006)
  [hep-ph/0605281].

\bibitem{Blanchet:2006be} 
  S.~Blanchet and P.~Di Bari,
  JCAP {\bf 0703}, 018 (2007)
  [hep-ph/0607330].

\bibitem{Pascoli:2006ie} 
  S.~Pascoli, S.~T.~Petcov and A.~Riotto,
  Nucl.\ Phys.\ B {\bf 774}, 1 (2007)
  [hep-ph/0611338].

\bibitem{Branco:2006hz} 
  G.~C.~Branco, A.~J.~Buras, S.~J\"ager, S.~Uhlig and A.~Weiler,
  JHEP {\bf 0709}, 004 (2007)
  [hep-ph/0609067].

\bibitem{De Simone:2006dd} 
  A.~De Simone and A.~Riotto,
  JCAP {\bf 0702}, 005 (2007)
  [hep-ph/0611357].

\bibitem{Sigl:1993} 
  G.~Sigl and G.~Raffelt,
  Nucl.\ Phys.\ B {\bf 406}, 423 (1993).

\bibitem{Akhmedov:1998qx} 
  E.~K.~Akhmedov, V.~A.~Rubakov and A.~Y.~Smirnov,
  Phys.\ Rev.\ Lett.\  {\bf 81}, 1359 (1998)
  [hep-ph/9803255].

\bibitem{Shaposhnikov:2008pf} 
  M.~Shaposhnikov,
  JHEP {\bf 0808}, 008 (2008)
  [arXiv:0804.4542 [hep-ph]].

\bibitem{Gagnon:2010kt} 
  J.~S.~Gagnon and M.~Shaposhnikov,
  Phys.\ Rev.\ D {\bf 83}, 065021 (2011)
  [arXiv:1012.1126 [hep-ph]].

\bibitem{Asaka:2011wq} 
  T.~Asaka, S.~Eijima and H.~Ishida,
  JCAP {\bf 1202}, 021 (2012)
  [arXiv:1112.5565 [hep-ph]].

\bibitem{Canetti:2012kh} 
  L.~Canetti, M.~Drewes, T.~Frossard and M.~Shaposhnikov,
  Phys.\ Rev.\ D {\bf 87}, 093006 (2013)
  [arXiv:1208.4607 [hep-ph]].

\bibitem{Shuve:2014zua} 
  B.~Shuve and I.~Yavin,
  Phys.\ Rev.\ D {\bf 89}, 075014 (2014)
  [arXiv:1401.2459 [hep-ph]].

\bibitem{Dev:2014laa}
  P.~S.~B.~Dev, P.~Millington, A.~Pilaftsis and D.~Teresi,
 Nucl.\ Phys.\ B {\bf 886}, 569 (2014) [arXiv: 1404.1003 [hep-ph]].

\bibitem{Dev:2014tpa} 
  P.~S.~B.~Dev, P.~Millington, A.~Pilaftsis and D.~Teresi,
  arXiv:1409.8263 [hep-ph].
  
\bibitem{Pais:1955sm}
  A.~Pais and O.~Piccioni,
  Phys.\ Rev.\  {\bf 100} (1955) 1487.
  
\bibitem{Kabir}  
	P.~K.~Kabir, \emph{The CP puzzle: strange decays of the neutral kaon}, Academic Press (1968). 
 
\bibitem{Boyanovsky:2006yg}
  D.~Boyanovsky and C.~M.~Ho,
  Phys.\ Rev.\ D {\bf 75} (2007) 085004
  [hep-ph/0610036].
  
\bibitem{Boyanovsky:2014uya}
  D.~Boyanovsky and L.~Lello,
  New J.\ Phys.\  {\bf 16} (2014) 063050
  [arXiv:1403.6366 [hep-ph]].
  
\bibitem{Boyanovsky:2014lqa}
  D.~Boyanovsky,
   Nucl.\ Phys.\ B {\bf 888}, 248 (2014) [arXiv:1406.5739 [hep-ph]].
  
\bibitem{Boyanovsky:2014una}
  D.~Boyanovsky,
  arXiv:1409.4265 [hep-ph].  
  
\bibitem{Prokopec:2003pj}
  T.~Prokopec, M.~G.~Schmidt and S.~Weinstock,
  Annals Phys.\  {\bf 314} (2004) 208
  [hep-ph/0312110].

\bibitem{Prokopec:2004ic}
  T.~Prokopec, M.~G.~Schmidt and S.~Weinstock,
  Annals Phys.\  {\bf 314} (2004) 267
  [hep-ph/0406140].

\bibitem{Baym:1961zz}
  G.~Baym and L.~P.~Kadanoff,
  Phys.\ Rev.\  {\bf 124} (1961) 287.

\bibitem{KB} 
  L. Kadanoff and G. Baym,
  {\it Quantum Statistical Mechanics}, Benjamin, New York (1962).

\bibitem{Blaizot:2001nr}
  J.~P.~Blaizot and E.~Iancu,
  Phys.\ Rept.\  {\bf 359} (2002) 355
  [hep-ph/0101103].
  
\bibitem{Berges:2004yj}
  J.~Berges,
  AIP Conf.\ Proc.\  {\bf 739} (2005) 3
  [hep-ph/0409233].

\bibitem{Cornwall:1974vz}
  J.~M.~Cornwall, R.~Jackiw and E.~Tomboulis,
  Phys.~Rev.~D {\bf 10}, 2428 (1974).

\bibitem{AmelinoCamelia:1992nc} 
  G.~Amelino-Camelia and S.~Y.~Pi,
  Phys.\ Rev.\ D {\bf 47}, 2356 (1993)
  [hep-ph/9211211].

\bibitem{Berges:2005hc} 
  J.~Berges, S.~Bors\'anyi, U.~Reinosa and J.~Serreau,
  Annals Phys.\  {\bf 320}, 344 (2005)
  [hep-ph/0503240].

\bibitem{Pilaftsis:2013xna}
  A.~Pilaftsis and D.~Teresi,
  Nucl.\ Phys.\ B {\bf 874} (2013) 2,  594
  [arXiv:1305.3221 [hep-ph]].

\bibitem{Schwinger:1961} 
  J.~S.~Schwinger,
  J.\ Math.\ Phys.\  {\bf 2}, 407 (1961).

\bibitem{Keldysh:1964} 
  L.~V.~Keldysh,
  Zh.\ Eksp.\ Teor.\ Fiz.\  {\bf 47}, 1515 (1964)
  [Sov.\ Phys.\ JETP {\bf 20}, 1018 (1965)].

\bibitem{Jordan:1986ug}
  R.~D.~Jordan,
  Phys.\ Rev.\ D {\bf 33} (1986) 444.

\bibitem{Calzetta:1986ey}
   E.~Calzetta and B.~L.~Hu,
   Phys.~Rev.~D {\bf 35}, 495 (1987).

\bibitem{Calzetta:1986cq}
  E.~Calzetta and B.~L.~Hu,
  Phys.~Rev.~D {\bf 37}, 2878 (1988).


%
%

\bibitem{Buchmuller:2000nd} 
  W.~Buchm\"{u}ller and S.~Fredenhagen,
  Phys.\ Lett.\ B {\bf 483}, 217 (2000)
  [hep-ph/0004145].

\bibitem{De Simone:2007rw} 
  A.~De Simone and A.~Riotto,
  JCAP {\bf 0708}, 002 (2007)
  [hep-ph/0703175].

\bibitem{De Simone:2007pa} 
  A.~De Simone and A.~Riotto,
  JCAP {\bf 0708}, 013 (2007)
  [arXiv:0705.2183 [hep-ph]].

\bibitem{Cirigliano:2007hb} 
  V.~Cirigliano, A.~De Simone, G.~Isidori, I.~Masina and A.~Riotto,
  JCAP {\bf 0801}, 004 (2008)
  [arXiv:0711.0778 [hep-ph]].

\bibitem{Anisimov:2008dz} 
  A.~Anisimov, W.~Buchm\"uller, M.~Drewes and S.~Mendizabal,
  Annals Phys.\  {\bf 324}, 1234 (2009)
  [arXiv:0812.1934 [hep-th]].

\bibitem{Garny:2009rv}
  M.~Garny, A.~Hohenegger, A.~Kartavtsev and M.~Lindner,
  Phys.\ Rev.\ D {\bf 80} (2009) 125027
  [arXiv:0909.1559 [hep-ph]].

\bibitem{Garny:2009qn} 
  M.~Garny, A.~Hohenegger, A.~Kartavtsev and M.~Lindner,
  Phys.\ Rev.\ D {\bf 81}, 085027 (2010)
  [arXiv:0911.4122 [hep-ph]].

\bibitem{Cirigliano:2009yt} 
  V.~Cirigliano, C.~Lee, M.~J.~Ramsey-Musolf and S.~Tulin,
  Phys.\ Rev.\ D {\bf 81}, 103503 (2010)
  [arXiv:0912.3523 [hep-ph]].

\bibitem{Anisimov:2010aq} 
  A.~Anisimov, W.~Buchm\"uller, M.~Drewes and S.~Mendizabal,
  Phys.\ Rev.\ Lett.\  {\bf 104}, 121102 (2010)
  [arXiv:1001.3856 [hep-ph]].

\bibitem{Garny:2010nj}
  M.~Garny, A.~Hohenegger and A.~Kartavtsev,
  Phys.\ Rev.\ D {\bf 81} (2010) 085028
  [arXiv:1002.0331 [hep-ph]].
  
\bibitem{Beneke:2010wd} 
  M.~Beneke, B.~Garbrecht, M.~Herranen and P.~Schwaller,
  Nucl.\ Phys.\ B {\bf 838}, 1 (2010)
  [arXiv:1002.1326 [hep-ph]].

\bibitem{Beneke:2010dz} 
  M.~Beneke, B.~Garbrecht, C.~Fidler, M.~Herranen and P.~Schwaller,
  Nucl.\ Phys.\ B {\bf 843}, 177 (2011)
  [arXiv:1007.4783 [hep-ph]].

\bibitem{Garbrecht:2010sz} 
  B.~Garbrecht,
  Nucl.\ Phys.\ B {\bf 847}, 350 (2011)
  [arXiv:1011.3122 [hep-ph]].

\bibitem{Anisimov:2010dk} 
  A.~Anisimov, W.~Buchm\"{u}ller, M.~Drewes and S.~Mendizabal,
  Annals Phys.\  {\bf 326}, 1998 (2011)
  [Erratum-ibid.\  {\bf 338}, 376 (2011)]
  [arXiv:1012.5821 [hep-ph]].

\bibitem{Garbrecht:2011aw} 
  B.~Garbrecht and M.~Herranen,
  Nucl.\ Phys.\ B {\bf 861}, 17 (2012)
  [arXiv:1112.5954 [hep-ph]].

\bibitem{Garny:2011hg} 
  M.~Garny, A.~Kartavtsev and A.~Hohenegger,
  Annals Phys.\  {\bf 328}, 26 (2013)
  [arXiv:1112.6428 [hep-ph]].

\bibitem{Drewes:2012ma}
  M.~Drewes and B.~Garbrecht,
  JHEP {\bf 1303} (2013) 096
  [arXiv:1206.5537 [hep-ph]].

\bibitem{Garbrecht:2012pq}
  B.~Garbrecht,
  Nucl.\ Phys.\ B {\bf 868} (2013) 557
  [arXiv:1210.0553 [hep-ph]].

\bibitem{Frossard:2012pc} 
  T.~Frossard, M.~Garny, A.~Hohenegger, A.~Kartavtsev and D.~Mitrouskas,
  Phys.\ Rev.\ D {\bf 87}, 085009 (2013)
  [arXiv:1211.2140 [hep-ph]].

\bibitem{Drewes:2013gca}
  M.~Drewes,
  Int.\ J.\ Mod.\ Phys.\ E {\bf 22} (2013) 1330019
  [arXiv:1303.6912 [hep-ph]].

\bibitem{Garbrecht:2013iga}
  B.~Garbrecht and M.~J.~Ramsey-Musolf,
  Nucl.\ Phys.\ B {\bf 882} (2014) 145
  [arXiv:1307.0524 [hep-ph]].

\bibitem{Hohenegger:2013zia}
  A.~Hohenegger and A.~Kartavtsev,
  arXiv:1309.1385 [hep-ph].

\bibitem{Iso:2013lba} 
  S.~Iso, K.~Shimada and M.~Yamanaka,
  arXiv:1312.7680 [hep-ph].

\bibitem{Iso:2014afa} 
  S.~Iso and K.~Shimada,
  JHEP {\bf 1408}, 043 (2014)
  [arXiv:1404.4816 [hep-ph]].
 
\bibitem{Hohenegger:2014cpa}
  A.~Hohenegger and A.~Kartavtsev,
  JHEP {\bf 1407} (2014) 130
  [arXiv:1404.5309 [hep-ph]].

\bibitem{Garbrecht:2014aga}
  B.~Garbrecht, F.~Gautier and J.~Klaric,
  JCAP {\bf 1409} (2014) 09,  033
  [arXiv:1406.4190 [hep-ph]].

\bibitem{Weldon:1991ek}
  H.~A.~Weldon,
  Phys.\ Rev.\ D {\bf 45} (1992) 352.

\bibitem{Altherr:1994fx}
  T.~Altherr and D.~Seibert,
  Phys.\ Lett.\ B {\bf 333} (1994) 149
  [hep-ph/9405396].

\bibitem{Altherr:1994jc}
  T.~Altherr,
  Phys.\ Lett.\ B {\bf 341} (1995) 325
  [hep-ph/9407249].

\bibitem{Bedaque:1994di}
  P.~F.~Bedaque,
  Phys.\ Lett.\ B {\bf 344} (1995) 23
  [hep-ph/9410415].

\bibitem{Dadic:1998yd}
  I.~Dad\'ic,
  Phys.\ Rev.\ D {\bf 59} (1999) 125012
  [hep-ph/9801399].

\bibitem{Greiner:1998ri}
  C.~Greiner and S.~Leupold,
  Eur.\ Phys.\ J.\ C {\bf 8} (1999) 517
  [hep-ph/9804239].

\bibitem{Garbrecht:2011xw}
  B.~Garbrecht and M.~Garny,
  Annals Phys.\  {\bf 327} (2012) 914
  [arXiv:1108.3688 [hep-ph]].

\bibitem{Winter:1986da}
  J.~Winter,
  Phys.\ Rev.\ D {\bf 32} (1985) 1871.

\bibitem{Berges:2005md}
  J.~Berges and S.~Bors\'anyi,
  Phys.\ Rev.\ D {\bf 74} (2006) 045022
  [hep-ph/0512155].

\bibitem{Garbrecht:2008cb}
  B.~Garbrecht and T.~Konstandin,
  Phys.\ Rev.\ D {\bf 79} (2009) 085003
  [arXiv:0810.4016 [hep-ph]].

\bibitem{Garny:2010nz}
  M.~Garny, A.~Hohenegger and A.~Kartavtsev,
  arXiv:1005.5385 [hep-ph].
  
\bibitem{Vlasenko:2013fja}
  A.~Vlasenko, G.~M.~Fuller and V.~Cirigliano,
  Phys.\ Rev.\ D {\bf 89} (2014) 105004
  [arXiv:1309.2628 [hep-ph]].

\bibitem{Lipavsky:1986zz}
  P.~Lipavsk\'y, V.~\v{S}pi\v{c}ka and B.~Velick\'y,
  Phys.\ Rev.\ B {\bf 34} (1986) 6933.

\bibitem{Bornath:1996zz}
  T.~Bornath, D.~Kremp, W.~D.~Kraeft and M.~Schlanges,
  Phys.\ Rev.\ E {\bf 54} (1996) 3274.

\bibitem{Herranen:2010mh}
  M.~Herranen, K.~Kainulainen and P.~M.~Rahkila,
  JHEP {\bf 1012} (2010) 072
  [arXiv:1006.1929 [hep-ph]].

\bibitem{Herranen:2011zg}
  M.~Herranen, K.~Kainulainen and P.~M.~Rahkila,
  JHEP {\bf 1202} (2012) 080
  [arXiv:1108.2371 [hep-ph]].

\bibitem{Fidler:2011yq}
  C.~Fidler, M.~Herranen, K.~Kainulainen and P.~M.~Rahkila,
  JHEP {\bf 1202} (2012) 065
  [arXiv:1108.2309 [hep-ph]].

\bibitem{Millington:2012pf}
  P.~Millington and A.~Pilaftsis,
  Phys.\ Rev.\ D {\bf 88} (2013) 8,  085009
  [arXiv:1211.3152 [hep-ph]].

\bibitem{Millington:2013isa}
  P.~Millington and A.~Pilaftsis,
  Phys.\ Lett.\ B {\bf 724} (2013) 56
  [arXiv:1304.7249 [hep-ph]].

\bibitem{Kolb:1980} 
  E.~W.~Kolb and S.~Wolfram,
  Nucl.\ Phys.\ B {\bf 172}, 224 (1980)
  [Erratum-ibid.\ B {\bf 195}, 542 (1982)].

\bibitem{Canetti:2014dka} 
  L.~Canetti, M.~Drewes and B.~Garbrecht,
  arXiv:1404.7114 [hep-ph].

\end{thebibliography}
\end{document}